\documentclass[aps,prd,twocolumn,superscriptaddress,showpacs]{revtex4}
\usepackage{graphicx,epsf,amssymb,amsmath,latexsym}
\newcommand{\cu}{\mbox{${\mathcal U}$}}
\newcommand{\be}{\begin{eqnarray}}
\newcommand{\ee}{\end{eqnarray}}

\renewcommand{\d}{\mbox{{\rm d}}}

\def \D {\mbox{D}}

\def \cu{{\cal U}}
\def \cq{ Q}
\def \cp{ \Pi}

\begin{document}
\begin{flushleft}
DAMTP-2004-115
\end{flushleft}
\title{Gravitational collapse and black hole evolution:
do holographic black holes eventually ``anti-evaporate''?}
\author{Roberto Casadio}
\email{casadio@bo.infn.it}
\affiliation{Dipartimento di Fisica, Universit\`a di Bologna and I.N.F.N.,
Sezione di Bologna, via Irnerio~46, 40126 Bologna, Italy}
\author{Cristiano Germani}
\email{C.Germani@damtp.cam.ac.uk}
\affiliation{D.A.M.T.P., Centre for Mathematical Sciences,
University of Cambridge, Wilberforce road, Cambridge CB3 0WA, England}
\begin{abstract}
We study the gravitational collapse of compact objects in the
Brane-World. We begin by arguing that the regularity of the
five-dimensional geodesics does not allow the energy-momentum tensor
of matter on the brane to have (step-like) discontinuities, which
are instead admitted in the four-dimensional General Relativistic
case, and compact sources must therefore have an atmosphere.
Under the simplifying assumption that matter is a spherically
symmetric cloud of dust without dissipation, we can find the
conditions for which the collapsing star generically
``evaporates'' and approaches the Hawking behavior as the
(apparent) horizon is being formed.
Subsequently, the apparent horizon evolves into the
atmosphere and the back-reaction on the brane metric reduces
the evaporation, which continues until the effective energy
of the star vanishes.
This occurs at a finite radius, and the star afterwards
re-expands and ``anti-evaporates''.
We clarify that the Israel junction conditions across the
brane (holographically related to the matter trace anomaly)
and the projection of the Weyl tensor on the brane
(holographically interpreted as the quantum back-reaction
on the brane metric) contribute to the total energy
as, respectively, an ``anti-evaporation'' and an
``evaporation'' term.
Concluding, we comment on the possible effects of dissipation
and obtain a new stringent bound for the brane tension.
\end{abstract}
\pacs{04.50.+h, 04.70.-s, 04.70.Dy}
\maketitle
\section{Introduction}
\label{intro}
It is well known that black holes are unstable in four (and higher)
dimensions because of the Hawking effect~\cite{hawking}, that is the
quantum mechanical production of particles in strong inhomogeneous
gravitational fields.
It is also well known that such an effect leads (and is deeply linked)
to the trace anomaly of the radiation field on the black hole
background~\cite{CF,birrell}.
However, for a complete description, the semiclassical
Einstein equations should be solved including the {\em back-reaction\/}
of the evaporation flux on the metric, which turns out to be an
extremely hard task (for a recent attempt to incorporate the effect
of the trace anomaly see Ref.~\cite{vilasi}).
\par
In the context of the Randall-Sundrum (RS) Brane-World (BW)
models~\cite{RS}, it was shown in Ref.~\cite{BGM} (see also
Ref.~\cite{kofinas} for some recent generalizations) that the collapse
of a homogeneous star leads to a non-static exterior, contrary to what
happens in four-dimensional General Relativity (GR), and a possible
exterior was later found which is radiative~\cite{dad}.
If one regards black holes as the natural end state of the collapse,
one may conclude that {\em classical\/} black holes in the BW should
suffer of the same problem as {\em semiclassical\/} black holes in GR:
no static configuration for their exterior might be allowed.
\par
In particular, it was shown in Ref~\cite{holo}, that all
known black hole-like metrics on the brane lead to Weyl anomalies
with a natural interpretation in the context of the holographic
analogy \cite{holography}.
Moreover, such anomalies could be related with an instability,
as those metrics do not seem to have the correct weak field
expansion in RS (for a discussion of this issue see
Ref.~\cite{gm}).
Further, forcing a static exterior, a trace anomaly outside a
homogenous and isotropic collapsing star appears which is of
the same form, but with opposite sign, as that of semiclassical
black holes.
This suggested the possibility that black hole metrics which solve
the bulk equations with brane boundary conditions, and whose
central singularities are located on the brane, genuinely
correspond to quantum corrected (semiclassical) black holes on the
brane~\cite{tanaka1,fabbri}, in the spirit of the holographic
principle~\cite{holography} and AdS/CFT conjecture~\cite{AdSCFT}.
\par
We recall that our Universe is a codimension one four-dimensional
hypersurface of vacuum energy density $\lambda$ in the BW scenario
of Ref.~\cite{RS}.
It is hence useful to introduce Gaussian normal coordinates
$x^A=(x^\mu,y)$, where $y$ is the extra-dimensional coordinate such
that the brane is located at $y=0$ (capitol letters run from
$0$~to~$4$ and Greek letters from $0$~to~$3$).
The five-dimensional metric can then be expanded near the brane
as~\cite{maart}
\be
g^5_{AB}&=&\left.g^5_{AB}\right|_{y=0}
+2\,\left.K_{AB}\right|_{y=0}\, y
+\pounds_{\hat n}\left.K_{AB}\right|_{y=0}\,y^2
\nonumber
\\
&&
+\ldots
\ ,
\label{g5}
\ee
where $K_{AB}$ is the extrinsic curvature of the brane,
and $\pounds_{\hat n}$ the Lie derivative along the unitary
four-vector $\hat n$ orthogonal to the brane.
We also recall that the junction conditions at the brane lead
to~\cite{shiromizu}
\be
K_{\mu\nu}\sim
T_{\mu\nu} -\frac{1}{3}\,\left(T-\lambda\right)\,g_{\mu\nu}
\ ,
\label{K}
\ee
where $T_{\mu\nu}$ is the stress tensor of the matter localized
on the brane, and~\cite{maart}
\be
\pounds_{\hat n} K_{\mu\nu}\sim {\cal E}_{\mu\nu}+f(T)_{\mu\nu}
\ ,
\label{f}
\ee
where ${\cal E}_{\mu\nu}$ is the projection of the Weyl tensor on the
brane and $f(T)_{\mu\nu}$ a tensor which depends on $T_{\mu\nu}$ and
$\lambda$.
\par
The junction conditions in GR~\cite{israel} allow (step-like)
discontinuities in the stress tensor (for example, across the
surface of a star) keeping the first and second fundamental
forms continuous.
For thin (Dirac $\delta$-like) surfaces, a step-like discontinuity
of the extrinsic curvature orthogonal to the surface is also
allowed as long as the metric remains continuous~\cite{israel}.
Since a brane in RS is itself a thin surface, it generates an
orthogonal discontinuity of the extrinsic curvature in five
dimensions as allowed by GR junction conditions.
However, any discontinuities in the matter stress tensor on the
brane would induce discontinuities in the extrinsic curvature
(\ref{K}) which are tangential to the brane and would therefore
appear in the five-dimensional metric (\ref{g5}).
Such discontinuities of the metric are not allowed by the
regularity of five-dimensional geodesics.
Moreover, because of the second order term in Eq.~(\ref{g5}) and
considering Eq.~(\ref{f}), we can not allow the projected Weyl
tensor to be discontinuous on the brane either~\footnote{
Eq.~(\ref{g5}) is a Taylor expansion (not a perturbation) performed
in Gaussian normal coordinates.
There is thus no bending of the brane which can make the
five-dimensional metric continuous when there are discontinuities
in the matter stress tensor and/or the projected Weyl tensor.}.
One can understand the above regularity requirement by considering
that, in a microscopic description of the BW, matter should be
smooth along the fifth dimension, yet localized on the brane
(say, within a width of order $\lambda^{-1/2}$, in units $G=c=1$~\cite{g2}).
In any such description, the continuity of five-dimensional
geodesics must then hold and, in order to build a physical model
of a star, one has to smooth both the matter stress tensor and the
projected Weyl tensor across the surface of the star along
the brane.
\par
We shall employ the effective four-dimensional (hydrodynamical)
equations of Refs.~\cite{maart,shiromizu} in our analysis.
In general, such equations cannot determine the brane metric
uniquely unless one also knows the bulk geometry.
However, if the system enjoys enough symmetries, the effective
four-dimensional equations are closed and can thus give some
insight about the bulk~\footnote{For examples of exact and
perturbative models see, respectively, Refs.~\cite{BGM}
and~\cite{bruni}.}.
In particular, we shall show that, under the simplifying
assumption that the heat flow is always negligible
(no dissipation), the knowledge of the full five-dimensional
dynamics of the most central region of the collapsing star
renders the whole system ``physically closed'' when the energy
density of the star is much smaller than the brane tension.
By physically closed we mean that the evolution of the system
is uniquely determined upon further requiring that the
four-dimensional metric become Minkowski in the limit of zero
energy density and at spatial infinity (asymptotic flatness).
Although considering a non-dissipative model appears
restrictive, we would like to remark that the same kind
of models in GR leads to paradigmatic examples of black hole
formation, beside the fact that this is the only case which
can be solved analytically.
\par
The Oppenheimer-Snyder (OS) model in GR~\cite{OS} yields the
simplest description of how black holes could form from
collapsing stars.
It has however been shown that this kind of model is not
viable in the RS scenario~\cite{BGM} since, if one forces
a static exterior outside homogeneous and isotropic stars,
BW effects produce an effective ``energy surplus'' which
is encoded by a positive curvature in such an exterior
and which cannot be generated by any bulk back-reaction.
Hence, we expect that this excess energy will be released
via a mechanism that leads to a loss of mass from the star.
Although the positive curvature in the exterior has no GR
description, it can also be obtained from quantum
computations~\cite{CF}~\footnote{This anomaly is related
to the scale invariance of a massless scalar field and is
historically called Weyl anomaly.} but with the
{\em opposite\/} sign.
The sign mismatch between classical and quantum results
might be reconciled on recalling that the classical anomaly
is due to an effective potential energy at the boundary of
the star which must be released in order to have an exterior
compatible with the junction conditions~\cite{phd}.
It might therefore be possible to change the sign of the
anomaly just considering that the energy surplus should be
converted into an effective negative flux of energy from
the boundary of the star.
\par
In order to do so, we shall employ a Tolman geometry~\cite{tolman}
for the brane star, as it is the only spherically symmetric metric
which does not allow dissipation of energy across the shells of the
collapsing star, and the Hawking radiation can then be interpreted
as the emission of gravitons into the bulk.
In fact, it turns out that the propagation of CFT modes in four
dimensions is consistently described by this mechanism according
to the AdS/CFT~correspondence~\cite{tanaka1}.
Moreover, we require the continuity of the Weyl and energy-momentum
tensors as discussed above, and we shall show that a Tolman brane
metric corresponds to a general five-dimensional diagonal metric
with spherically symmetric slices~\footnote{We wish to thank
Akihiro~Ishibashi for pointing this out to us.}.
\par
We shall divide the domain of the star into three regions:
I)~the ``core'', where most of the energy of the star is
concentrated;
II)~a ``transition region'', which connects the core with
a tail and finally
III)~a ``tail'', where the energy density approaches zero.
The tail and transition region together form an ``atmosphere''
of the sort that is usually employed in numerical simulations of
the gravitational collapse (see, e.g.~Ref.~\cite{rezzolla}).
The core is taken homogeneous and isotropic (OS-like) for
several reasons.
Firstly, in order to consider a minimal modification with respect
to the OS model in GR.
Secondly, the OS core corresponds to an exact five-dimensional
solution~\cite{langlois} and reproduces the correct Weyl anomaly
of quantum field theory on the Schwarzschild
background~\cite{CF,birrell}, thus making the holographic
interpretation clearer.
Our main results will then be that, in this case,
the total energy of the system is conserved~\footnote{We
note in passing that this is the main assumption in the
microcanonical treatment of the black hole evaporation (see,
e.g.~Refs.~\cite{micro}).} and that the collapsing star
``evaporates'' until the core experiences a ``rebound'' in
the high energy regime (when its energy density is comparable
with $\lambda$), after which the whole system
``anti-evaporates''.
Moreover, we can find a range of parameters for which the
minimum radius of the collapsing core is larger than the AdS
length (which sets the scale of Quantum Gravity in the BW),
thus further supporting the qualitative behavior we obtain.
\par
In Section~\ref{holo}, we shall briefly review a simple
holographic interpretation of the Hawking radiation in the BW.
In Section~\ref{coll}, we shall build a physical model of the
Tolman type that converges to the OS model in the GR limit and
show that BW corrections lead to an emission of energy from
the star and a bounce of the core (see also
Appendix~\ref{App_T}).
In Section~\ref{BH}, we shall analyze in details how the
(apparent) horizon forms and physical quantities related to it,
such as the outer trace anomaly, which will then be interpreted
in terms of the BW corrections coming from the brane junction
conditions.
In section \ref{heat} we shall discuss the possible effects
of dissipation.
We shall finally comment on our results in Section~\ref{conc}.
\par
In the following, we shall use geometrical units with $G=c=1$
and mostly positive metric signature.
\section{Simple holographic picture}
\label{holo}
Before trying to ``cure'' the Weyl anomaly, let us have a closer
look at the features of the effective energy surplus in the
exterior of the star discovered in Ref.~\cite{BGM}.
We first note that, as pointed out by different
authors~\cite{holo,fabbri,phd},
the energy surplus reproduces the absolute value of the quantum
Weyl anomaly computed on a Schwarzschild background \cite{CF},
which is the unique exterior of a spherical star in GR.
More precisely, in static coordinates one has, outside of the star,
\be
R^\mu_{\ \mu}=\frac{9}{2\,\pi\,\lambda}\,\frac{M^2}{R^6}
\ ,
\ee
where $R^\mu_{\ \mu}$ is the Ricci scalar, $M$ the physical mass of
the star~\footnote{We will discuss later on the meaning of
``physical mass''.}, and $R$ the Schwarzschild radial coordinate.
As remarked in Ref.~\cite{fabbri}, the holographic interpretation
for such a contribution cannot be of the exact AdS/CFT kind,
because both classical black holes on the brane and semiclassical
black holes in GR correspond to strong deviation from AdS and CFT
(see also Ref.~\cite{holo}).
We shall indeed show that we cannot reproduce the evaporation
process if the bulk is simply AdS (i.e.~with zero Weyl tensor).
\par
As a first step, we shall show that, if a black hole is formed
from matter collapsing in the BW, the area of its horizon
(to first order in $\lambda^{-1}$ and for a short time after its
formation) follows the evaporation law for semiclassical black
holes~\cite{hawking}.
In particular, we will see that the horizon evaporates provided
the Weyl contribution is dominant, and we may therefore assert that
the BW collapse gives, to first order in $\lambda^{-1}$ and
for some five-dimensional geometries, a good description of the first
order quantum processes related to it.
Let us also note that quantum calculations in Refs.~\cite{hawking,CF}
are performed in adiabatic approximation, that is, in some sense,
to first order in the back-reaction parameter of the quantum theory.
\par
Following Ref.~\cite{BGM}, a unique static geometry which matches
a collapsing homogeneous and isotropic cloud of dust, has
a Schwarzschild-like metric with mass function
\be
M=M_{\rm S}+\frac{1}{\lambda}\,m(R)
\ ,
\label{Mholo}
\ee
where $M_{\rm S}$ is the usual ADM contribution
(see Section~\ref{coll} for more details) and
\be
m(R)=\frac{3\,M_{\rm S}^2}{8\,\pi\,R^3}
-\frac{9\,\mu}{32\,\pi\,R}
\ ,
\ee
where, in a cosmological background, the constant $\mu$ is related
with the mass of a black hole sitting in the bulk~\cite{bh} and
we set the effective four-dimensional cosmological constant to zero
(since we are just interested in BW effects on asymptotically
flat branes).
\par
We denote with $R_0=R_0(\tau)$ the radius of the collapsing object
which depends on the proper time $\tau$.
The geodesic equation of motion in the Schwarzschild-like space-time
determines $R_0$ according to~\footnote{We shall use the notation
$\dot f=\partial_\tau f$ and $f'=\partial_r f$ when it is not
confusing.}
\be
\dot R_0^2=\frac{2\,M}{R_0}
=\frac{2\,M_{\rm S}}{R_0}
+\frac{3}{4\,\pi\,\lambda\,R_0^2}\,\left(\frac{M_{\rm S}^2}{R_0^2}
-\frac{3}{4}\,\mu\right)
\ ,
\ee
in which we have selected the case corresponding to zero initial
velocity for infinite initial radius.
We can now see how the mass function is changing when the
surface of the star crosses its own horizon, that is
at the time $\tau_{\rm H}$ when
$R_0(\tau_{\rm H})\equiv R_{\rm H}=2\,M_{\rm H}\equiv
2\,M(\tau_{\rm H})=
2\,M_{\rm S}+2\,m(R_{\rm H})/\lambda$~\footnote{Whereas
$R_0$ is the physical radius of a particular comoving
trajectory of the collapsing fluid, the function
$2\,M_{\rm H}(\tau)$ represents the time-evolution of the
horizon which does not occur at fixed comoving radial
coordinate (see Section~\ref{hor}).
We must hence stress that the subscript H in this Section
merely indicates that a given function of $\tau$ is evaluated
at the time $\tau_{\rm H}$.}
and $\dot R_{\rm H}=-1$.
Let us define the surface area of the evolving
``apparent horizon" as $A_{\rm AH}=16\,\pi\,M^2(\tau)$,
for which Eq.~(\ref{Mholo}) gives
\be
\dot A_{\rm AH}=\frac{9}{4\,\lambda}\,
\left[\mu-\left(\frac{M_{\rm S}}{M}\right)^2\right]\,
\frac{\dot R}{M}
\ .
\ee
Considering that $M_S/M\simeq 1$ to first order in
$\lambda^{-1}$, at the time $\tau=\tau_{\rm H}$, we then have
\begin{subequations}
\be
\dot A_{\rm AH}\simeq
-\frac{9}{4\,\lambda}\,\frac{\mu-1}{M_{\rm H}}
\ ,
\label{A}
\ee
or
\be
\dot M_{\rm H} \simeq -\frac{9}{128\,\pi\,\lambda}\,
\frac{\mu-1}{M_{\rm H}^2}
\ .
\label{OSfl}
\ee
\end{subequations}
The collapse therefore leads to a negative flux
of energy when the boundary of the star approaches its horizon,
as expected for the Hawking evaporation, {\em if\/} $\mu>1$.
\par
Since a positive $\mu$ generally corresponds to a reinforcement of
the localization of gravity in RS~\cite{maa2}, we can assert that
an OS region mainly evaporates into gravitational waves propagating
on the brane.
We shall however see that, for a consistent model of collapsing
star with continuous density, the sign of the Weyl energy changes
from the interior to the exterior of the star, so that the
evaporation actually ejects energy off the brane via gravitational
waves (as suggested in Ref.~\cite{tanaka1}).
\par
So far, we have not considered any back-reaction on the brane metric,
and the same flux (\ref{OSfl}) will reasonably be seen by a distant
observer for whom $\partial_\tau$ asymptotically becomes a time-like
Killing vector.
However, the surplus energy must be released, since no BW or GR
model can explain the Weyl anomaly, and this directly implies that
Eq.~(\ref{OSfl}) probably holds only for a short time about the
formation of the horizon, as suggested in \cite{BGM}.
We shall indeed show that this is the case.
\section{Gravitational collapse on the brane}
\label{coll}
In this section we will study a continuous model for the
gravitational collapse.
In order to see the difference with respect to the OS-like
model studied in Ref.~\cite{BGM}, we consider a Tolman-like
model with a central OS core.
The star is therefore described as a cloud of dust with falling off
continuous density and no sharp boundary.
The classical four-dimensional behavior will be recovered in the
limit of negligible star density (with respect to the brane
vacuum energy density $\lambda$).
\subsection{General framework}
Following Ref.~\cite{maart}, we can rewrite the BW effective
four-dimensional Einstein equations with vanishing cosmological
constant on the brane as
\be
G_{\mu\nu}=8\,\pi\,T_{\mu\nu}^{\rm eff}
\ .
\label{eq:effective}
\ee
Here we have
\be
T_{\mu\nu}^{\rm eff}=
\rho^{\rm eff}\,u_\mu\,u_\nu
+ p^{\rm eff}\,h_{\mu\nu}
+q^{\rm eff}_{(\mu}\,u_{\nu)}
+\pi^{\rm eff}_{\mu\nu}
\ ,
\label{decT}
\ee
where $u^\mu$ is the unit four-velocity of matter
($u^\mu u_\mu=-1$), $h_{\mu\nu}$ the space-like metric
that projects orthogonally to $u^\mu$
($h_{\mu\nu}=g_{\mu\nu}+u_\mu u_\nu$) and $\pi^{\rm eff}_{\mu\nu}$
an anisotropic tensor.
\par
For an isotropic perfect fluid, BW corrections to GR are
described by the effective quantities \cite{maart}
\begin{subequations}
\be
\rho^{\rm eff} &=&
\rho\,\left(1+\frac{\rho}{2\,\lambda}+\frac{\cal U}{\rho}\right)
\label{reff}
\\
\nonumber
\\
p^{\rm eff } &=& p  + \frac{\rho}{2\,\lambda}
\left(2\,p+\rho\right)
+\frac{\cal U}{3}
\label{peff}
\\
\nonumber
\\
q^{\rm eff }_\mu &=& Q_\mu
\\
\nonumber
\\
\pi^{\rm eff }_{\mu\nu} &=& \Pi_{\mu\nu}
\ ,
\label{e:pressure2}
\ee
\end{subequations}
where $\rho$ and $p$ are the (``bare'') energy density and
pressure of matter.
We also employed the following decomposition of the projection
of the Weyl tensor on the brane
\be
-{\frac{1}{8\pi}}\,{\cal E}_{\mu\nu}&=&
\cu\,\left(u_\mu\,u_\nu+\frac{1}{3}\,h_{\mu\nu}\right)
\nonumber
\\
&&
+{\cq_\mu}\,u_{\nu} + {\cq_\nu}\,u_{\mu}+\cp_{\mu\nu}
\ ,
\ee
corresponding to an effective ``dark'' radiation on the brane with
energy density $\cal U$, pressure ${\cal U}/3$, momentum density
$Q_\mu$ and anisotropic stress ${\Pi}_{\mu\nu}$.
Note that non-local bulk effects can contribute to effective
imperfect fluid terms even when brane matter is a perfect fluid.
\par
Bianchi identities supplied by the junction conditions produce
two kinds of conservation equations~\cite{maart}:
\begin{enumerate}
\item
Local conservation equations (LCE):
\begin{subequations}
\be
&&
\dot{\rho}+\Theta\,\left(\rho+p\right)=0
\label{pc1}
\\
\nonumber
\\
&&
\D_a p+\left(\rho+p\right)\,A_a =0
\ ;
\label{pc2}
\ee
\end{subequations}
\item
Non-local conservation equations (NLCE's):
\begin{subequations}
\be
&&
\!\!\!\!\!\!
\dot{\cu}+\frac{4}{3}\,\Theta\,{\cu}+\D^ a\cq_a+2\,A^a\,\cq_a
+\sigma^{ab}\,\cp_{ab}=0
\label{pc1'}
\\
\nonumber
\\
&&
\!\!\!\!\!\!
\dot\cq_{a}+{\frac{4}{3}}\,\Theta\,\cq_a +{\frac{1}{3}}\,\D_a{\cu}
+\frac{4}{3}\,{\cu}\,A_a +\D^b\cp_{ab}
\nonumber
\\
&&
+A^ b\,\cp_{ab}
+\sigma_{a}^{\ b}\,\cq_b-\omega_a^{\ b}\,\cq_b
=-\frac{\rho+p}{\lambda}\,\D_a\rho
\ ,
\label{pc2'}
\ee
\end{subequations}
\end{enumerate}
where $D_a$ is the spatially projected derivative
(defined by
$D_a S^{b...}{}_{...c}=h^e{}_a h^b{}_f...h^g{}_c \nabla_e S^{f...}{}_{...g}$
for $a=1,\ 2,\ 3$),
$\Theta=\nabla^\alpha u_\alpha$ the volume expansion,
$\dot S^{a...}{}_{...b}=u^\alpha\,\nabla_\alpha S^{a...}{}_{...b}$
the proper time derivative, $A_a=\dot u_a$ the acceleration,
$\sigma_{ab}=D_{(a} u_{b)}-(\Theta/3)\,h_{ab}$
the (traceless) shear, and $\omega_{ab}=-D_{[a}u_{b]}$
the vorticity.
\subsection{Spherically symmetric dust}
\label{ssd}
For the case with zero pressure ($p=0$), that is dust on the brane,
the quantities in Eqs.~(\ref{reff}), (\ref{peff}) and
(\ref{e:pressure2}) reduce to
\begin{subequations}
\be
\rho^{\rm eff}&=&\rho\,\left(1+\frac{\rho}{2\,\lambda}\right)
+{\cal U}
\\
\nonumber
\\
p^{\rm eff }&=&\frac{\rho^2}{2\,\lambda}+\frac{\cal U}{3}
\\
\nonumber
\\
\pi^{\rm eff}_{\mu\nu} &=& \Pi_{\mu\nu}
\ .
\ee
\end{subequations}
Provided the matter density $\rho$ does not vanish in the region of
interest, one can use comoving coordinates in which
$u^\alpha=(-1,0,0,0)$.
\par
In the following, we will only consider the class of
five-dimensional metrics which are diagonal (sufficiently
close to the brane at $y=0$) and spherically symmetric
on the brane.
In Gaussian normal coordinates, one can always write
a bulk metric which is spherically symmetric on the brane
as
\be
\!\!\!
\d s^2&\!\!=\!\!&-N^2(\tau,r,y)\,\d \tau^2
+A^2(\tau,r,y)\,\d r^2
\nonumber
\\
&&
+2\,B(\tau,r,y)\,\d t\,\d r
+R^2(\tau,r,y)\,\d\Omega^2
\nonumber
\\
&&
+\d y^2
\ .
\label{bulk_m}
\ee
Upon using the restricted freedom to change the
four-dimensional coordinates on the brane,
one can always set $B(\tau,r,0^+)=0$~\cite{LL},
so that the brane metric reads
\be
\left.\d s^2\right|_{y=0^+}
&=&
-N^2(\tau,r,0^+)\,\d\tau^2
+A^2(\tau,r,0^+)\,\d r^2
\nonumber
\\
&&
+R^2(\tau,r,0^+)\,\d\Omega^2
\ .
\ee
Since we just consider dust as brane matter,
from the junction conditions at the brane~\cite{maart},
we also obtain
\be
0=K_{\tau r}^+(\tau,r)
\equiv\left. \frac{1}{2}\,\frac{\partial g_{\tau r}}{\partial y}\right|_{y=0^+}
=\left.\frac{\partial B}{\partial y}\right|_{y=0^+}=0
\ .
\ee
Using the above result together with the bulk symmetry
$Z_2$ with respect to the brane, we have
$B(\tau,r,y)=y^2\,\left[V(\tau,r)+{\mathcal O}(y)\right]$.
Since the Weyl energy flux is related to $B$ by
\be
Q_a\sim
\left.
\frac{\partial^2 B}{\partial y^2}
\right|_{y=0^+}
\ee
one finds that $Q_a$ vanishes if $V(\tau,r)=0$,
which is in fact what we are assuming.
The coefficient $g_{\tau r}$ then vanishes fast enough
on the brane so that, from the five-dimensional
Einstein equations
\be
^{(5)}G_{AB}=-\Lambda\,g_{AB}
\ ,
\ee
in the limit $y\to 0^+$, one obtains the
condition~\footnote{We thank Christophe~Galfard
for discussions about this point.}
\be
0=\!\!\left.^{(5)}\!G_{\tau r}\right|_{y=0^+}
\!\!\!\!=
\left.
\frac{2}{NA}\,\left(
\frac{\dot A}{A}\,\frac{R'}{R}
+\frac{\dot R}{R}\,\frac{N'}{N}
-\frac{\dot R'}{R}
\right)\right|_{y=0^+}
\!\!\!
,
\label{five}
\ee
where a prime denotes $\partial_r$ and a dot $\partial_\tau$.
Since our matter is pressureless, we can work in the proper time
gauge $N(\tau,r,0^+)=1$~\cite{LL} and, using the residual
gauge freedom in defining the radial coordinate $r$,
we obtain
\be
A(\tau,r,0^+)=R'(\tau,r,0^+)
\ .
\ee
This relation implies a Tolman geometry on the brane~\footnote{We
consider here only the case corresponding to a
sphere of dust collapsing from infinite radius with vanishing
initial velocity, which is the spatially flat case.
However, since this is just a kinematical detail, spatially curved
cases should be qualitatively the same.}
\be
\d s^2=-\d\tau^2+\left(R'\right)^2\,\d r^2+R^2\,\d\Omega^2
\ ,
\label{tolman}
\ee
where $R=R(\tau,r)$ is a (generally non-separable) function of $\tau$
and $r$ such that $4\,\pi\,R^2(\tau,r)$ equals the surface area of
the shell comoving with dust particles located at the coordinate
position $r$ at the proper time $\tau$.
\par
With the above symmetries, the vorticity, the acceleration and
the Weyl energy flux vanish, $\omega_a=A_a=Q_a=0$, and we obtain the
simplified LCE
\be
\partial_\tau{\rho}+\Theta\,\rho=0
\label{drho}
\ ,
\ee
and NLCE's
\begin{subequations}
\be
&&
\partial_\tau{\cu}+{\frac{4}{3}}\,\Theta\,{\cu}
+\sigma^{ab}\,\Pi_{ab}=0
\\
\nonumber
\\
&&
{\frac{1}{3}}\,D_a{\cu}+D^b\Pi_{ab}=-\frac{\rho}{\lambda}\,D_a\rho
\ .
\ee
\end{subequations}
The volume expansion is also easily computed as
\be
\Theta=\partial_\tau\left[\ln \left(R^2\,\partial_r R\right)\right]
=\frac{\partial_\tau\partial_r\left(R^3\right)}
{\partial_r\left(R^3\right)}
\ ,
\ee
and for the shear one finds
\be
\sigma_{ab}=\frac{1}{2}\,\partial_\tau h_{ab}-\frac{\Theta}{3}\,h_{ab}
\ ,
\ee
where $h_{ab}=g_{ab}$ is the spatial part of the metric
(\ref{tolman}).
\par
By symmetry, we expect that the anisotropic pressure tensor is
diagonal and isotropic in the angular directions.
Moreover, considering that $\Pi^\alpha{}_\alpha=0$ we have, in such
adapted coordinates,
\be
\Pi^a_{\ b}
=\mbox{diag}\,
\left(\frac{2}{3}\,\Pi,-\frac{1}{3}\,\Pi,-\frac{1}{3}\,\Pi\right)
\ ,
\ee
and
\be
\sigma^{ab}\,\Pi_{ab}=\frac{1}{2}\,\partial_\tau g_{ab}\,\Pi^{ab}
=-\frac{2}{3}\,\Pi\,\left(\frac{\partial_\tau R}{R}-
\frac{\partial_\tau\partial_r R}{\partial_r R}
\right)
\ ,
\ee
which vanishes in the OS background (homogeneous and isotropic
space-time) for which
\be
R(\tau,r)=g(r)\,X(\tau)
\ .
\label{R_OS}
\ee
We then see that the NLCE's become
\begin{subequations}
\be
&&\!\!\!\!\!\!\!\!
\dot \cu+\frac{4}{3}\,\left[
\frac{\dot R}{R}\,\left(
2\,\cu-\frac{\Pi}{2}\right)+\frac{\dot R'}{R'}\,
\left(\cu+\frac{\Pi}{2}\right)
\right]
=0
\label{first}
\\
\nonumber
\\
&&\!\!\!\!\!\!\!\!
\frac{1}{3}\,\cu'
+\frac{2}{3}\,\left[\Pi'+3\,\frac{R'}{R}\,\Pi\right]
=-\frac{\rho}{\lambda}\,\rho'
\ .
\label{second}
\ee
\end{subequations}
The system of NLCE's is in general not closed, since we do not have
an evolution equation for $\Pi$.
However, for a sufficiently large physical radius $R$, the knowledge of
$\Pi$ in an extended spatial region together with the asymptotic
flatness and the continuity of ${\cal E}_{\mu\nu}$ make that
system closed.
Let us remark that this also happens in the cosmological perturbative
scenario in which one considers large-scale evolution of the Weyl
tensor~\cite{bruni}.
\par
The LCE~(\ref{drho}) integrated over the spatial volume
$\sqrt{h}\,\d r\,\d\theta\,\d\phi
=\sin\theta/3\,\partial_r(R^3)\,\d r\,\d\theta\,\d\phi$ implies
\be
\partial_\tau m_\rho=0
\ ,
\label{dM0}
\ee
where we have introduced the ``bare'' mass function
\be
m_\rho(r)\equiv \frac{4\,\pi}{3}\,\int^{r}_0
\rho(\tau,x)\,\partial_x\left(R^3(\tau,x)\right)\,\d x
\ ,
\label{M_0}
\ee
or, equivalently,
\be
\rho(\tau,r)=\frac{m_\rho'}{4\,\pi\,R^2\,R'}
\ .
\label{rho}
\ee
The meaning of Eq.~(\ref{dM0}) is that, since we have chosen
a comoving reference frame and $p=0$, the ``bare'' energy
contained within a sphere of fixed coordinate radius $r$ cannot change
in time, although the physical radius $R(\tau,r)$ of such a sphere
decreases during the collapse.
\par
We can now consider the $(\tau,\tau)$ Einstein equation,
\be
G^\tau_{\ \tau}=-\frac{(\dot R^2\,R)'}{R^2\,R'}
=-8\,\pi\,\rho^{\rm eff}
\ ,
\label{G}
\ee
which yields the equation of motion
\be
\dot R^2(\tau,r)=\frac{2\,M(\tau,r)+F(\tau)}{R(\tau,r)}
\ ,
\ee
where we have introduced the ``effective'' mass
\be
M(\tau,r)=\frac{4\,\pi}{3}\,\int^{r}_0
\rho^{\rm eff}(\tau,x)\,\partial_x \left(R^3(\tau,x)\right)\,
\d x
\ .
\label{Meffg}
\ee
Since we want a flat brane for $M=0$ [moreover, the center
of the star is at rest, $\dot R(\tau,0)=0$], it must be
$F(\tau)=0$ and we finally obtain
\be
\dot R^2(\tau,r)=\frac{2\,M(\tau,r)}{R(\tau,r)}
\ .
\label{R}
\ee
\par
Let us note that the effective mass is not constant in general.
In fact,
\be
\!\!\!\!\!\!
\dot M(\tau,r)&=&\frac{4\,\pi}{3}\,\partial_\tau
\int^{r}_0
\rho^{\rm eff}\,\partial_x(R^3)\,\d x
\nonumber
\\
&=&\frac{4\,\pi}{3}\,\int^{r}_0
\partial_\tau
\left[\left(\frac{\rho^2}{2\,\lambda}+\cu\right)\,
\partial_x \left(R^3\right)\right]\,\d x
\ .
\ee
For the particular case $\Pi=0$, one then obtains
\be
\dot M(\tau,r)
=-\frac{4\,\pi}{3}\,\int^{r}_0
\left(\frac{\rho^2}{2\,\lambda}+\frac{\cu}{3}\right)\,
\partial_x\partial_\tau \left(R^3\right)\,\d x
\ ,
\label{dMeff}
\ee
where we have used both the LCE and the first NLCE.
\par
A very important result which follows from the LCE and NLCE's is
that, {\em if the brane metric is asymptotically flat, the anisotropic
stress $\Pi\not=0$ whenever $\dot\rho'\neq 0$\/}.
We can prove it by showing that $\Pi=0$ is not compatible with
asymptotic flatness and the LCE and NLCE's.
On combining Eq.~(\ref{first}) with Eq.~(\ref{drho}) for $\Pi=0$,
we obtain
\be
\cu=U(r)\,\rho^{4/3}
\ ,
\label{327}
\ee
where $U(r)$ is a time-independent integration function.
From Eq.~(\ref{second}) with $\Pi=0$, one instead obtains
\be
\cu=-\frac{3\,\rho^{2}}{2\,\lambda}+F(\tau)
\ ,
\label{328}
\ee
$F(\tau)$ being a spatially-constant integration function.
Asymptotic flatness requires that
\be
\lim_{r\to\infty}\cu(\tau,r)=
\lim_{r\to\infty}\rho(\tau,r)=0
\ ,
\ \ \ \
\forall\,\tau
\ ,
\ee
which implies $F(\tau)=0$.
On now combining Eq.~(\ref{327}) with Eq.~(\ref{328}),
we get the relation
\be
U(r)=-\frac{3\,\rho^{2/3}}{2\,\lambda}
\ ,
\ee
which obviously contradicts the assumption $\dot\rho'\neq 0$.
This implies that, for a continuous distribution of dust for which
$\dot\rho'\not=0$, one must have $\Pi\not=0$.
This result supports the holographic interpretation as it resembles
very much a property of the renormalized quantum stress tensor on
the Schwarzschild background~\cite{CF}.
\subsection{The model}
\begin{figure}[t]
\centering
\epsfxsize=2.8in
\epsfbox{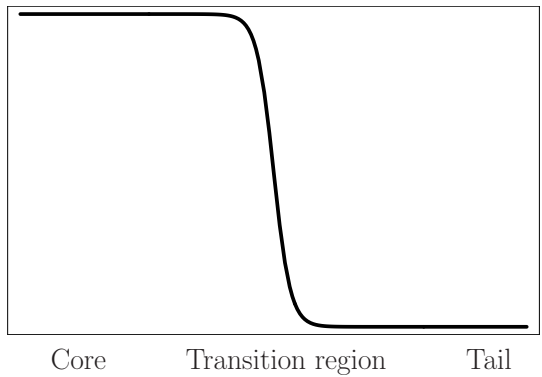}
\caption{Density profile.
\label{figura}}
\end{figure}

As mentioned before, we shall divide the star in three
regions~\footnote{The boundaries between any two regions
are considered as limits.} (see Fig.~\ref{figura} for a qualitative
picture):
\begin{description}
\item[I$\ \ $]: the ``core'' ($0\le r< r_0$), where $\rho'= 0$ and
one has an OS \cite{OS} behavior;
\item[II$\ $]: the ``transition region'' ($r_0< r<r_{\rm s}$), with Tolman
\cite{tolman} behavior due to $\rho\,\rho'$ being non-negligible;
\item[III]: the ``tail'' ($r> r_{\rm s}$), with $\rho\,\rho'\simeq 0$.
\end{description}
Moreover, we define the dimensionless parameter
\be
\epsilon\equiv {\rho_0/ \lambda}
\ ,
\ee
where $\rho_0\equiv \rho(\tau=0,r=0)$ is the initial core density.
Such a parameter is assumed small, since the system is initially
in a low energy regime (from the BW point of view) and relevant
quantities can thus be expanded in powers of $\epsilon$ for
sufficiently short times (or sufficiently large distance from the
core).
\par
A basic feature of both Tolman and OS models in four-dimensional GR is
that the bare mass function at fixed comoving radius is constant
in time and remains well defined during all the collapse.
Therefore, dust shells of different comoving radius move along
geodesics solely determined by the inner geometry and reach the
central singularity ($R=0$) at increasing proper times (Tolman model)
or at the same proper time (OS model).
In the former case one can have an enlarging apparent
horizon~\footnote{We briefly recall that, depending on the initial
conditions, the apparent horizon might start forming after the central
singularity, thus leaving it naked for some time.}, while in the latter
just an event horizon forms at the star surface~\cite{LL}.
\par
In the BW, the role of the bare mass is taken by the
effective mass $M$ of Eq.~(\ref{Meffg}), which will be
shown to diverge whenever $R\to 0$, thus making the whole
four-dimensional space-time singular.
To avoid this case, which is mathematically admissible but
physically unlikely, one has to include a sufficiently negative
contribution to the mass coming from the projected Weyl tensor.
As we discussed in Section~\ref{holo}, this will generate an
Hawking flux near the forming horizon, and we shall further show
that the effective mass completely evaporates at a finite star
radius, after which the collapse changes to a re-expansion
(or ``anti-evaporation'' process).
This case of BW collapse and rebound cannot be related to the GR
behavior perturbatively (in $\epsilon\sim\lambda^{-1}$),
since none of the shells reach $R=0$, but we incidentally note that
it seems in agreement with the uncertainty principle of quantum
mechanics~\footnote{It has been speculated
that classical BW equations reproduce four-dimensional quantum
equations in Ref.~\cite{wesson}.}.
In fact, a ``bounce'' in the trajectories of the collapsing matter
caused by quantum gravitational fluctuations had already been found
in an improved semiclassical analysis of the OS model~\cite{impBO}.
\subsubsection{The core}
\label{s_core}
We first recall that the bulk solution which corresponds to the OS
core of the star is perfectly regular in five dimensions
far from the space-time singularity~\cite{langlois}.
Further, since $\rho'= 0$, the system of relevant equations is now
closed.
In fact, we have $\Pi=0$, and the NLCE's reduce to the one
equation
\be
\dot{\cu}+\frac{4}{3}\,\Theta\,{\cu}=0
\ ,
\ee
which is solved by
\be
\cu=-\frac{27\,\mu\,r^4\,\epsilon}{128\,\pi^2\,r_0^4\,\rho_0\,R^4}
\ ,
\label{Ocu}
\ee
where $\mu$ is a constant.
\par
The physical radius $R$ can in general be written in the factorized
form (\ref{R_OS}) and the coordinate $r$ can be so chosen that
\be
g(r)=\left(\frac{9}{2}\,M_{\rm S}\right)^{1/3}\,\frac{r}{r_0}
\ ,
\label{g(r)}
\ee
in which $M_{\rm S}$ is the total bare mass of the OS core.
The effective mass~(\ref{Meffg}) is then given inside the core by
\be
\!\!\!\!\!\!\!
M(\tau,r)&\!\!=\!\!&M_{\rm S}\left(\frac{r}{r_0}\right)^3
\nonumber
\\
\!\!\!\!
&&
+\frac{9\,\epsilon}{32\,\pi\,\rho_0}
\left(\frac{r}{r_0}\right)^4\left[
\frac{(2\,M_{\rm S})^2}{3\,R^3}\left(\frac{r}{r_0}\right)^2
-\frac{\mu}{R}\right]
\ ,
\label{Mos}
\ee
where the first term in the r.h.s.~is the usual bare mass and the
remainder represents the BW correction.
\par
The above effective mass would diverge for $R\to 0$ (this also occurs
for a general Tolman core, see~Appendix~\ref{App_T}).
The point $R=0$ is the usual central singularity, which is harmless
(at least when covered by an horizon) in four-dimensional
GR, since the bare mass is constant and finite.
In the present case, however, the diverging effective energy makes
the whole space-time singular.
In order to see this, let $T$ be the proper time at which the OS
core hits the singularity.
From the equation of motion (\ref{R}) one has
\be
\left.R\,\dot R^2\right|_{r>r_0}
=2\,M(T,r_0)+8\,\pi\,\int_{r_0}^r \rho^{\rm eff}\,R^2\,R'\,\d x
\ .
\ee
Since $M(\tau\to T,r_0)\to\infty$, either the second term in the r.h.s.~is
finite and the total effective mass diverges at any $r>r_0$, thus
making the whole exterior singular, or it equals $-M(T,r_0)+f(r)$,
with $f(r)$ a regular function, in order to compensate for the
diverging core energy.
In the latter case, the Weyl energy becomes everywhere
infinitely large and negative and, since $G^{a}_{\ a}\sim\cu$,
the whole four-dimensional Einstein tensor is singular.
Although such singular evolutions appear mathematically allowed
by the equations, in the following we shall not consider them
since, from the BW point of view, either they predict a
catastrophic end of the Universe induced by astrophysical events
or, more reasonably, they suggest that a more fundamental model
must be used.
However, in the latter case we expect for large black holes
that a huge energy flux would be emitted towards infinity
well before the OS boundary approaches the Planck length.
This, of course, would be ruled out by astronomical observations.
From the holographic perspective we are interested in here,
only the non-singular solution is relevant.
In fact, it is only in this case that the star continuously
``evaporates'' (before the bouncing) by emitting a Hawking-like
energy flux at the moment when the OS horizon forms, as we shall
see later.
Moreover the bouncing solution seems to be compatible with some
proposal for the quantum black hole formation~\cite{Haw,impBO}.
\par
In order to avoid the singular cases, one must have $\mu$ positive
and large enough so that each shell will bounce back after reaching
a minimum radius where the corresponding effective mass
vanishes~\footnote{This peculiarity was already noted for a
pure Weyl collapse in the BW in Ref.~\cite{BGM}.}.
The Weyl tensor, holographically interpreted as the quantum
back-reaction on the brane metric (see~\cite{phd} and
References therein), then contributes the ``evaporation'' term
proportional to $\mu$ in Eq.~(\ref{Mos}), which dominates at
relatively low energies;
whereas, the BW correction to the matter stress tensor,
holographically interpreted as the matter quantum trace
anomaly~\cite{tanaka1}, yields the ``anti-evaporating'' term
proportional to $M_S^2$ in Eq.~(\ref{Mos}), which increases
with the energy.
\par
Upon inserting the effective mass (\ref{Mos}) in the equation
of motion (\ref{R}), one obtains an equation for $X(\tau)$,
\be
\dot X^2&=&
\frac{4}{9\,X}
+\frac{\epsilon}{27\,\pi\,\rho_0\,X^{4}}
-\frac{6^{1/3}\,\epsilon\,\mu}{24\,\pi\,\rho_0\,M_{\rm S}^{4/3}\,X^{2}}
\nonumber
\\
&\equiv&
-V(X)
\ ,
\label{Xeq}
\ee
in which there is no dependence on $r$.
This shows that the system remains ``rigid'' through the bounce:
no shell crossing occurs and all shells reach their minimum
radius at the same proper time.
Like for the classical OS model, it is thus sufficient to
consider the evolution of the core surface at $r=r_0$ and
we correspondingly define $M_0(\tau)=M(\tau,r_0)$ and
$R_0(\tau)=R(\tau,r_0)$.
\par
\begin{figure}[t]
\centering {\raisebox{3.5cm}{$V$}}
\epsfxsize=2.8in
\epsfbox{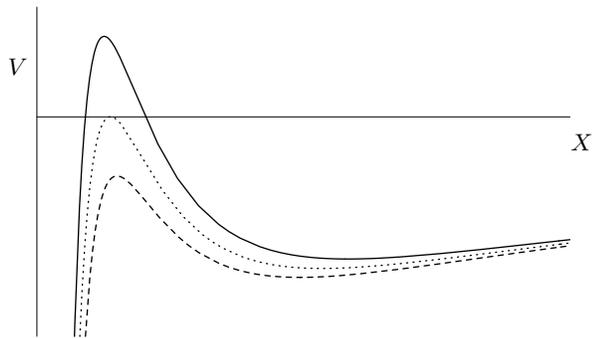}
{\raisebox{2.5cm}{$X$}}
\caption{Qualitative behavior of the shells and core potential  for
$\mu>\mu_{\rm c}$ (solid line) and $\mu<\mu_{\rm c}$ (dashed line).
For $\mu=\mu_{\rm c}$ (dotted line), the peak of $V$ equals the shells
energy $E=0$.
\label{V(X)}}
\end{figure}
\begin{figure}[ht]
\centering{
\epsfxsize=2.8in
\epsfbox{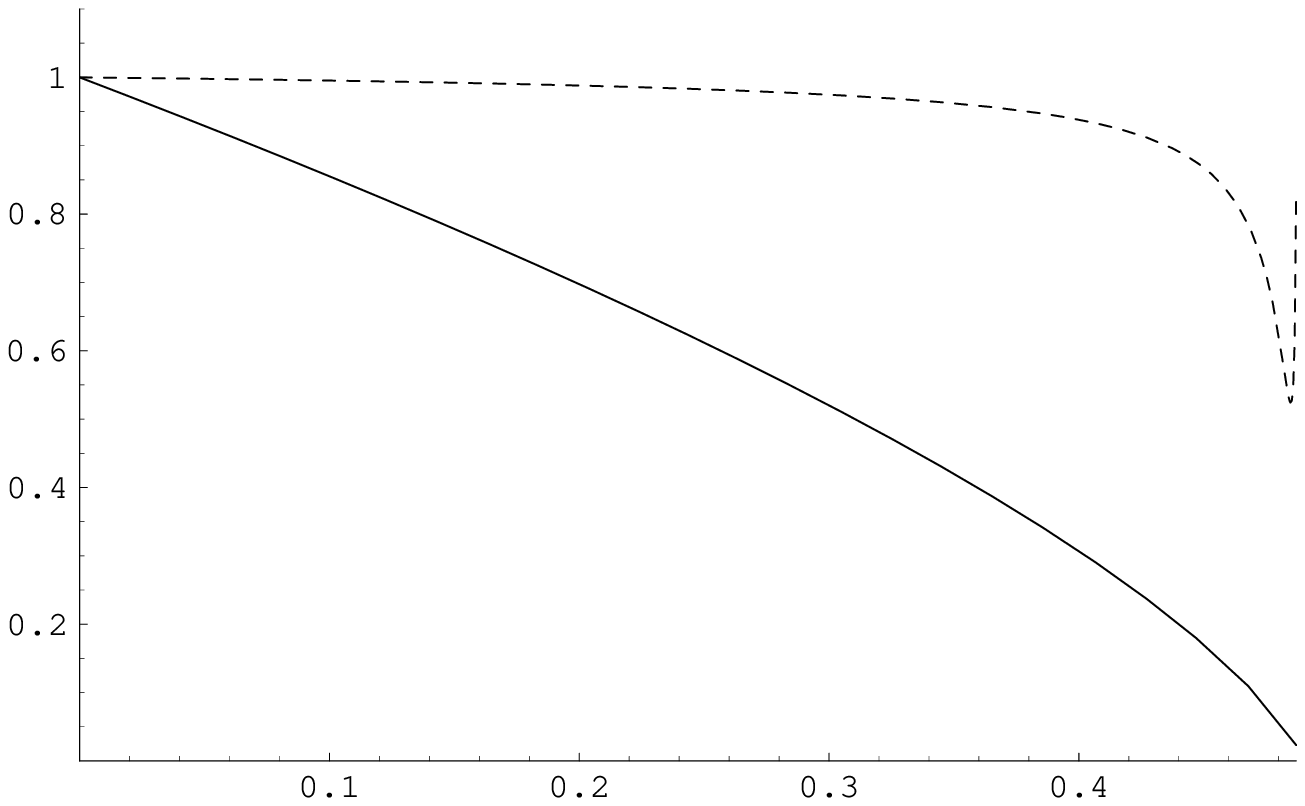}
$\tau$
\\
(a)
\\
\epsfxsize=2.8in
\epsfbox{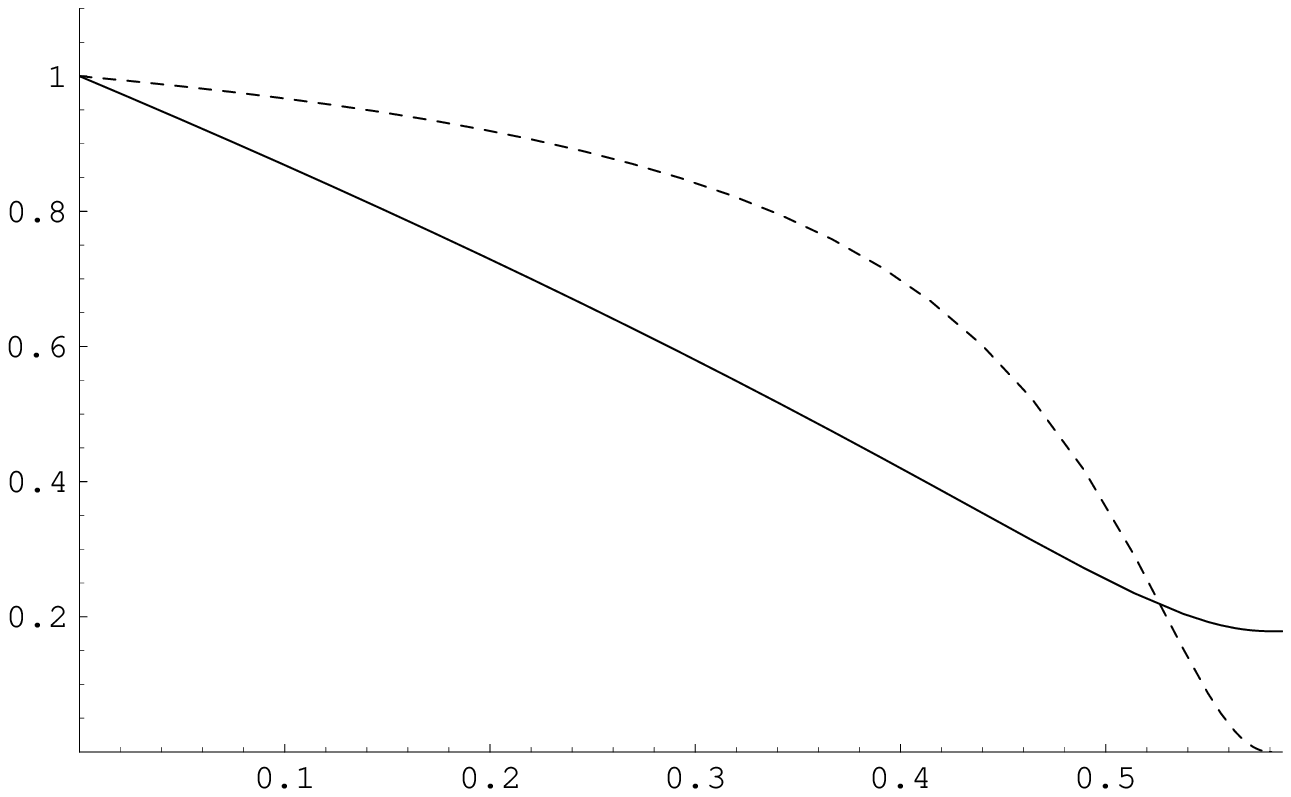}
$\tau$
\\
(b)}
\caption{Typical evolutions of the core radius $R_0(\tau)/R_0(0)$
(solid line) and effective mass $M_0(\tau)/M_0(0)$ (dashed line) for
$\mu<\mu_{\rm c}$ (a) and for $\mu>\mu_{\rm c}$ (b).
Units are arbitrary.
\label{mu<mu_c}}
\end{figure}
From the form of the potential $V$ in Eq.~(\ref{Xeq})
(see~also Fig.~\ref{V(X)}), one can see that the term proportional
to $\mu$ behaves as a repulsive (angular momentum-like) force and
the bounce occurs whenever there is a positive peak (since
the energy of collapsing shells $E=0$ for our choice of
initial conditions).
There will in general exist a critical value
$\mu_{\rm c}=\mu_{\rm c}(M_{\rm S},\rho_0,\epsilon)$ such that
one has the bounce for $\mu>\mu_{\rm c}$,
otherwise $R_0\to 0$ and $M_0$ diverges.
For $\mu=\mu_{\rm c}$ the two turning points of the potential
coincide and the shells would take an infinite proper time to reach
the minimum radius (of course, this would only occur if one neglected
any perturbations, and we shall not further consider this special case).
In Fig.~\ref{mu<mu_c} we display a typical trajectory of $R_0$,
along with the corresponding time evolution of the core effective
mass $M_0$, for $\mu<\mu_{\rm c}$ in panel~(a) and for
$\mu>\mu_{\rm c}$ in~(b).
In the latter case, after the core surface has reached the point
of zero effective mass, it will bounce back transforming the
whole collapse into an ``explosion'', which evolves as in~(b)
with the time reversed.
Although we are not able to describe the dynamics of the
atmosphere at very high energies (e.g.~around the bounce)
by means of our perturbative analysis,
on considering that the continuity of the total energy-momentum
tensor would be spoiled if the shells crossed~\footnote{Moreover,
shell crossings would prevent us from using a comoving frame
and a metric of the form~(\ref{tolman}).
We also note that, in four-dimensional GR, there are claims that
shell crossings lead to the formation of naked singularities
(see, e.g.~Ref.~\cite{naked}).},
one finds that all the shells (both in the core and the
atmosphere) must begin to re-expand.
Because of energy conservation at infinity, this ``reversal
of motion'' in the atmosphere would generate an instantaneous
distributional (Dirac $\delta$-like) term in the Ricci scalar,
as it was also found in a semiclassical treatment of bouncing solutions~\cite{Russo}.
Such a singularity in turn means that a detailed description
of the collisions between matter shells at fixed $r$ inside the
atmosphere and with the core must be taken into account at that
point.
A microscopic description of the shells goes however beyond
the scope of the present paper and we just wish to make a
remark.
In practice, during the bounce the collisionless description
of dust must be relaxed by introducing an effective short
distance potential which results in an effective equation of
state for the atmosphere.
Since our system is non-dissipative by construction, the
scatterings should be completely elastic and the equation of
state of the polytropic type (see, e.g.~Ref.~\cite{rezzolla}).
Of course, this property can be viewed as an artifact
of our simplified model, whereas in a more realistic situation
some energy will be dissipated from both the core and the
atmosphere, as we shall discuss in Section~\ref{heat}.
\par
The turning points of $R_0$ can be found analytically
by solving the cubic equation $V=0$, but their expression
is rather cumbersome.
It is instead easy to determine $\mu_{\rm c}$ exactly by observing
that the peak $V=0$ is located at
$X=X_{\rm p}\equiv (4\,M_{\rm S}/3)^{2/3}\,\mu^{-1/2}$
for some values of $M_{\rm S}$ and $\mu$, and, in general,
\be
V(X_{\rm p})=
\sqrt{\mu}\,
\frac{3\,\epsilon\,\mu^{3/2}-32\,\pi\,\rho_0\,M_{\rm S}^2}
{48\,6^{1/3}\,\pi\,\rho_0\,M_{\rm S}^{8/3}}
\ .
\ee
Hence, $V$ has two positive zeros if and only if $V(X_{\rm p})>0$,
that is when
\be
\mu>\left(\frac{32\,\pi\,\rho_0}{3\,\epsilon}\,M_{\rm S}^2\right)^{2/3}
\equiv \mu_{\rm c}
\ .
\label{2_0}
\ee
\par
We remark that the bouncing is a high energy effect compared to
$\epsilon$ (since $\mu_{\rm c}\sim \epsilon^{-2/3}$), whereas the
evaporation also occurs at low energy.
In fact, the core effective mass is given by
\be
M_0(\tau)=
M_{\rm S}+\frac{9\,\epsilon}{32\,\pi\,\rho_0}\,
\left[\frac{(2\,M_{\rm S})^2}{3\,R_0^3}-\frac{\mu}{R_0}\right]
\label{bouncing}
\ee
and its time derivative is
\be
\dot M_0(\tau)=
-\frac{9\,\epsilon}{32\,\pi\,\rho_0}\,
\left[\left(\frac{2\,M_{\rm S}}{R_0}\right)^2
-\mu\right]\,\frac{\dot R_0}{R_0^2}
\ .
\ee
Recalling that $\dot R_0<0$ during the collapse and that
$R_0(0)\gg 2\,M_{\rm S}$ (the initial core radius must
be outside the GR horizon),
we see that the evaporation sets out at the beginning of the collapse
when the star is still in the low energy regime.
Moreover, thanks to the condition~(\ref{2_0}), one can easily show
that $\dot M_0<0$, at least until the radius bounces back.
\par
In particular, the minimal radius $R_{min}$ is given by
\be
R_{\rm min}>
\lambda^{-1/2}\,\left(
\frac{M_{\rm S}}{\lambda^{-1/2}}\right)^{1/3}
\ .
\label{sim}
\ee
The holographic description is expected to hold only if the AdS
length $\ell\sim \lambda^{-1/2}$ is much shorter then the
typical lengths of the process we are considering~\cite{Porrati}.
The shortest length in our system is obviously given by
$R_{\rm min}$, for which we should therefore have
\be
\frac{R_{\rm min}}{\lambda^{-1/2}}\gg 1
\ .
\label{hc}
\ee
From Eq.~(\ref{sim}), we thus need
\be
M_{\rm S}\gg\lambda^{-1/2}
\ ,
\label{fur}
\ee
that is, the Schwarzschild radius of the star must be much
larger than the AdS length as one would have expected.
Furthermore, from Eq.~(\ref{fur}) it also follows that
$\mu_c\gg 1$.
\par
In light of this remark, in the following we will study the
evolution of the whole system to first order in $\epsilon$,
to which we have
\be
\dot M_0(\tau)=\mp
\frac{9\,\epsilon}{32\,\pi\,\rho_0}\,
\left[\mu-\frac{4\,M_{\rm S}^2}{R_0^2(\tau)}\right]\,
\frac{\sqrt{2\,M_{\rm S}}}{R_0^{5/2}(\tau)}
\ ,
\label{dm0finale}
\ee
where the minus sign ($\dot M_0<0$) holds during the collapse
and the plus sign ($\dot M_0>0$) after the bounce, and $R_0(\tau)$
can be determined to zeroth order in $\epsilon$.
\par
From now on, we shall just analyze the collapse, since the
explosion is the time reversal of the latter in our case where
there is no dissipation.
Since it is the core which first enters a high energy regime,
we can obtain a (rather conservative) estimate for the error
made if we truncate expressions to first order in $\epsilon$
by comparing $\dot M_0$ to first and second order by means
of the function~\footnote{We remind the
reader that $R_0$ and $M_0$ depend on $\epsilon$.}
\be
\Delta(\tau)\equiv
\left|
\frac{\left.\partial^2_\epsilon\dot M_0\right|_{\epsilon=0}\epsilon^2}
{\left.\partial_\epsilon\dot M_0\right|_{\epsilon=0}\epsilon}
\right|
\ ,
\label{errorF}
\ee
and consider that our approximation is good if $\Delta\lesssim 0.5$.
The analytic expression of $\Delta$ is extremely involved and we just
show a few plots in Appendix~\ref{AppDelta}, from which the
dependence on $M_{\rm S}$ and $\mu$ can be qualitatively
inferred.
\par
A physical upper bound on $|\mu|$ can be placed
by considering that the BW correction to the core bare mass for
astrophysical objects must be much smaller then the bare mass,
\be
\left|M_0-M_{\rm S}\right|\ll M_{\rm S}
\ ,
\label{upper_mu}
\ee
(at least) until the core approaches the GR horizon
($R_0\sim 2\,M_{\rm S}$), and Eq.~(\ref{bouncing}) then yields
\be
|\mu|\ll \frac{64\,\pi\,\rho_0}{9\,\epsilon}\,M_{\rm S}^2
\equiv \mu_{\rm a}
\ .
\label{error_mu}
\ee
For astrophysical objects one also expects $\lambda\,M_{\rm S}^2\gg 1$,
so that $\mu_{\rm a}\gg \mu_{\rm c}$.
Moreover, the limit (\ref{error_mu}) assures that the formation
of the OS (apparent) horizon, occurs before the bouncing.
However, this upper bound cannot likely be used for small black holes
for which we expect a strong Hawking evaporation even at the
formation of the first horizon.
\subsubsection{The transition region}
For $r_0<r< r_{\rm s}$, we are in the transition between two
regions of almost constant density.
Since in the GR model $\rho=0$ for $r>r_0$,
the energy outside the OS star is entirely a BW correction.
The density therefore must decrease rapidly from a value which is
of order $\epsilon^0$ to a value of order $\epsilon$.
This can be formalized as
\be
m_\rho(r;r_0)\equiv
\frac{4\,\pi}{3}\,\int_{r_0}^{r}\rho\,\left(R^3\right)'\,\d x
=O\left(\epsilon\right)
\ ,
\label{int}
\ee
where $r_0$ is again the border between the regions~I and~II and
the LCE as usual guarantees that $m_\rho$ remains constant.
Moreover, since the transition is overall a BW effect, we can take
\be
r_{\rm s}-r_0=O(\lambda^{-1/2})=O(\epsilon)
\ ,
\label{rs-r0}
\ee
and therefore
\be
R(\tau,r)-R_0(\tau)=O(\epsilon)
\ .
\label{R=R0}
\ee
Since $\cu=O(\epsilon)$, we also have that
\be
m_\cu(\tau,r;r_0)=
\frac{4\,\pi}{3}\,
\int^{r}_{r_0}\cu\,(R^3)'\,\d x=O\left(\epsilon^{2}\right)
\ ,
\label{intU}
\ee
for $r_0<r<r_{\rm s}$, and the contribution of $\cu$ to the
effective mass in region~II can be neglected.
Although in the transition region we have no control on the
projected Weyl tensor, we can still regard the system as closed
since the Weyl contribution does not affect the evolution at the
level of approximation we are considering.
Combining these results, we obtain that, to first order in $\epsilon$,
the effective mass is given by
\be
M(\tau,r)&\simeq& M_0(\tau)+m_\rho(r;r_0)
+m_\cu(\tau,r;r_0)
\nonumber
\\
&\simeq&
M_0(\tau)+m_\rho(r;r_0)
\ .
\label{MII}
\ee
This implies that, to first order in $\epsilon$,
\be
\dot M(\tau,r)\simeq \dot M_0(\tau)
\ .
\label{mII}
\ee
for $r_0<r<r_{\rm s}$, in agreement with the condition (\ref{rs-r0}),
and we can conclude that, since $\dot M<0$ at $r=r_0$, it will remain
negative (and substantially unaffected) throughout the border
of the transition region $r=r_{\rm s}$.
\subsubsection{The tail}
As in the transition region, $\rho=O(\epsilon)$ for $r_{\rm s}<r$,
and $m_\rho(r;r_{\rm s})=O(\epsilon)$.
Furthermore, we can now consider that in this regime
$\rho'\,\rho/\lambda=O\left(\epsilon^2\right)$,
so that bulk gravitons are decoupled from brane matter.
The Weyl contribution is however of the same order,
\be
m_\cu(\tau,r;r_{\rm s})=
\frac{4\,\pi}{3}\,\int_{r_{\rm s}}^r
\cu\,\left(R^3\right)'\,\d x
=O(\epsilon)
\ .
\label{m_U_s}
\ee
\par
We recall that the effective Einstein equations imply
that the Ricci scalar $R^\mu_{\ \mu}=-8\,\pi\,T^{\rm eff}$,
that is
\be
R^\mu_{\ \mu}=
\frac{3}{\left(R^3\right)'}\,\partial_r\left[R\,\partial_\tau^2R^2\right]
=8\,\pi\,\left(\rho-\frac{\rho^2}{\lambda}\right)
\ .
\label{Ricci}
\ee
Upon integrating over regions~II and III and taking into account
Eq.~(\ref{int}), we thus obtain, to first order in $\epsilon$,
\be
\left.R\,\partial_{\tau}^2\,
R^{2\phantom{\frac{I}{I}}}\!\!\!\right|_{r_0}^{r}
\simeq 2\,m_\rho(r;r_0)
\ ,
\ee
for $r_0<r$.
From the equation of motion (\ref{R}), the above relation yields
\be
\left.\frac{R}{\dot R}\,\dot M\right|_{r_{\rm s}}^{r}&=&
\left.R\,\partial_{\tau}^2\,
R^{2\phantom{\frac{I}{I}}}\!\!\!\right|_{r_{\rm s}}^{r}
-2\,m_\rho(r;r_{\rm s})
-2\,m_\cu(\tau,r;r_{\rm s})
\nonumber
\\
&\simeq&
-2\,m_\cu(\tau,r;r_{\rm s})
\ ,
\label{1}
\ee
now for $r_{\rm s}<r$.
On further considering Eq.~(\ref{intU}), we obtain
\be
\left.\frac{R}{\dot R}\,\dot M\right|_{r_0}^{r}
&\simeq&
\left.\frac{R}{\dot R}\,\dot M\right|_{r_{\rm s}}^{r}
\simeq
-2\,m_\cu(\tau,r;r_{\rm s})
\ .
\label{2}
\ee
Since $\dot M_0=O(\epsilon)$, we can use the zeroth order
equation of motion for the shells at fixed $r>r_{\rm s}$,
\be
\dot R^2(\tau,r)\simeq \frac{2\,M_{\rm S}}{R(\tau,r)}
\ ,
\label{motion0}
\ee
where $M_{\rm S}$ is again the total bare mass of the OS core.
Solutions to the above equations can be written as
\be
R(\tau,r)=\left(\frac{9}{2}\,M_{\rm S}\right)^{1/3}\,
\left[f(r)+T-\tau\right]^{2/3}
\ ,
\label{Rmot0}
\ee
where the function $f(r)$ is monotonically increasing in $r$ and
such that $R(\tau,r)$ is continuous across $r=r_{\rm s}$.
There is no loss of generality in assuming that $f(r)=r-c$ with
$c$ a constant, since changing $f$ is tantamount to redefining
the coordinate $r$.
In particular, on considering Eq.~(\ref{R=R0}), we can set $c=r_0$
to zeroth order in $\epsilon$ (for a discussion of $T$,
see Appendix~\ref{AppDelta}).
\par
One can now prove a general result  which holds irrespective of
the specific solutions for $\cu$ and $\Pi$.
Since, for $r>r_{\rm s}$,
\be
M(\tau,r)\simeq M_0(\tau)+m_\rho(r;r_0)+m_\cu(\tau,r;r_{\rm s})
\ ,
\label{MIII}
\ee
one has that
\be
\dot M(\tau,r)\simeq\dot M_0(\tau)+\dot{m}_\cu(\tau,r;r_{\rm s})
\ ,
\label{dMtl}
\ee
and, for any given $R=R(\tau,r)$, Eq.~(\ref{2}) becomes a differential
equation for $m_\cu=m_\cu(\tau,r;r_{\rm s})$, whose form further
simplifies on taking into account the approximation~(\ref{motion0}),
\be
\dot{m}_\cu+2\,\sqrt{\frac{2\,M_{\rm S}}{R^3}}\,m_\cu=
\dot M_0\,\left[\left(\frac{R_0}{R}\right)^{3/2}-1\right]
\ .
\label{dmeq}
\ee
Since $R(\tau,r)>R_0(\tau)$ for $r>r_0$, Eq. (\ref{dmeq}) implies that
$m_\cu$ cannot remain zero in the tail (note that the r.h.s.~is positive
for $\dot M_0<0$).
\par
In order to proceed, we now assume that:
\begin{description}
\item[(i) ]
{\em $\lim_{r\to\infty} R(\tau,r)=\infty$}~\footnote{This means
that the comoving reference frame extends all over the brane or,
at least, over a range much larger than the zeroth order size of the
star $R_0(0)$.}, and
\item[(ii)]
{\em the effective mass be finite at spatial infinity\/},
\be
\lim_{r\to\infty} M(\tau,r)<\infty
\ ,
\ \ \ \forall\,\tau>0
\ .
\ee
\end{description}
Since $M_0(\tau)$ always remains finite if $\mu>\mu_{\rm c}$
(i.e.~when there is a bounce),
and the bare mass of the tail is finite (and small) by construction,
this implies that
\be
\lim_{r\to\infty} m_\cu(\tau,r;r_{\rm s})<\infty
\ ,
\ \ \ \forall\,\tau>0
\ .
\ee
Since asymptotic flatness ensures that at large distance from
the core the low energy approximation holds, we can take the limit
$r\to\infty$ (equivalent to $R\to\infty$ at fixed time)
in Eq.~(\ref{dmeq}) and finally obtain
\be
\lim_{r\to\infty} \dot{m}_\cu(\tau,r;r_{\rm s})=-\dot M_0(\tau)
\ ,
\ \ \ \forall\,\tau>0
\ ,
\ee
or, from Eq.~(\ref{dMtl}),
\be
\lim_{r\to\infty} \dot M(\tau,r)=0
\ ,
\ \ \ \forall\,\tau>0
\ .
\label{dMinfty}
\ee
To summarize, we have shown that {\em if the total effective mass at spatial
infinity is finite at the initial time $\tau=0$, it will always remain
constant\/} (for a bouncing core evolution with $\mu>\mu_{\rm c}$), so that
{\em the total effective mass of the collapsing dust star is actually
conserved\/}.
\begin{figure}[t]
\centering
{\raisebox{2.3cm}{$m_\cu$}}
\epsfxsize=2.8in
\epsfbox{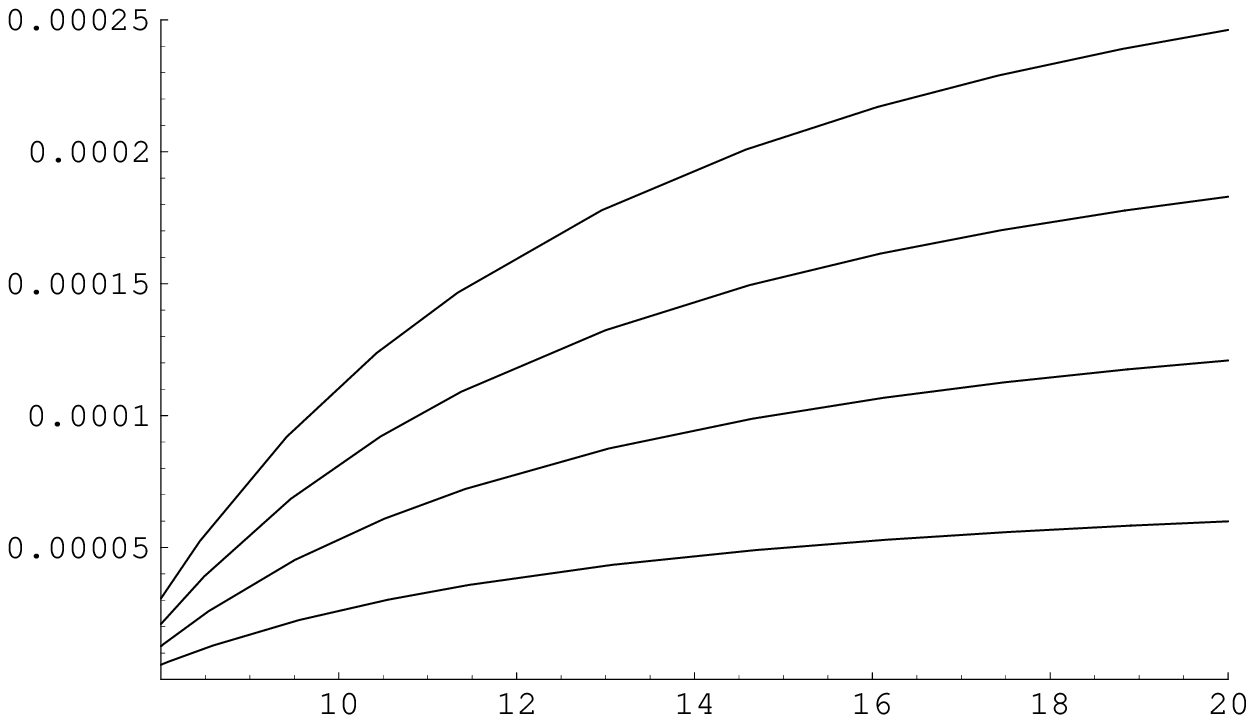}
\\
$\ \ \ \ \ \ \ \ \ \ \ \ \ \ \ \ \ \ \ \ \ \ \ \ \ \ \ \
\ \ \ \ \ \ \ \ \ \ \ \ \ \ \ \ \ \ \ \ \ \ \ \ \ \ \ \ \ R/M_{\rm S}$
\caption{Typical contribution of $\cu$ to the effective mass at
five subsequent times (from horizontal axis for $\tau=0$ to
upper curve for $\tau=T/20$) for $\epsilon=10^{-4}$,
$M_{\rm S}=\rho_0=1$, $T=10$ and $\mu=5000>4824=\mu_{\rm c}$.
\label{U}}
\end{figure}
\begin{figure}[t]
\centering
$\ \ \ \ \ \ \ \ \ \ \ \ \ \ \ \ \ \ \ \ \ \ \ \ \ \ \ \
\ \ \ \ \ \ \ \ \ \ \ \ \ \ \ \ \ \ \ \ \ \ \ \ \ \ \ \ \ R/M_{\rm S}$
\\
{\raisebox{2.3cm}{$\dot M$}}
\epsfxsize=2.8in
\epsfbox{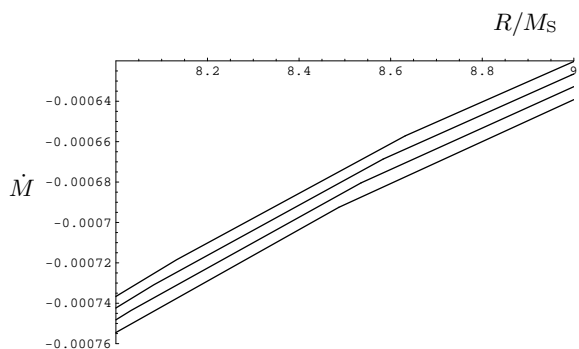}
\caption{Time derivative of the effective mass $M(\tau,r)$
versus the physical radius $R(\tau,r)$ for four values of
$\tau=T/100,\ldots, T/20$ (upper curve to lower curve) and
the same parameters used in Fig.~\ref{U}.
\label{dMext0}}
\end{figure}
\par
Eq.~(\ref{dmeq}) can be solved analytically with a generic initial
condition $m_\cu(0,r;r_{\rm s})$. It is particularly interesting
to consider the case $m_\cu(0,r;r_{\rm s})=0$ so that there is
initially no energy stored in the Weyl component $\cu$ (the bulk
metric is AdS at low energies). In this case, using
Eq.~(\ref{dmeq}), we have that $m_\cu>0$ during the collapse. This
yields the curves of Fig.~\ref{U} and the time derivative of the
effective mass of Fig.~\ref{dMext0}, in which we set
$\epsilon=10^{-4}$, $M_{\rm S}=\rho_0=1$, $T=10$ and
$\mu=5000>4824=\mu_{\rm c}$, but we note that different values of
these parameters do not change the qualitative behavior of $m_\cu$
and $\dot M$. In the following, we shall use these values of the
parameters for all the numerical computations and plots. They can
in fact be considered as the case of a small black hole for
which, however, the Holographic bound is satisfied.
The only purpose of the plots is to show more clearly the
qualitative behavior of the processes involved as well as to
support our perturbative expansion for any physical interesting
cases.
\par
Since $m_\cu(\tau,r;r_{\rm s})$ increases monotonically
in $r$ (starting from zero at $r=r_{\rm s}$) and,
because of Eq.~(\ref{dMinfty}), the effective mass of the star
outside the core
will decrease by releasing
gravitational waves off the brane and into the bulk~\cite{maa2}.
After the bounce, since the core will eventually re-enter a
low energy regime in which $\dot M_0$ is the same as the one
found before with opposite time evolution, we expect that $m_\cu$
will also evolve backwards so as to ensure the general condition
(\ref{dMinfty}).
At a time equal to twice the time of the bounce, we should
therefore have $m_\cu=0$, corresponding to the initial state
with zero Weyl energy.
This behavior is related to the non-dissipative nature of
our model, and we shall later discuss the possible effects
of dissipation.
\section{Black hole formation and evaporation}
\label{BH}
We now proceed to analyze the model developed in the previous
Section near the (forming) horizon.
\subsection{Horizons}
\label{hor}
We recall that shells of constant $r$ reach the (apparent) horizon
at the time $\tau=\tau_{\rm H}(r)$ when
\be
\dot R(\tau_{\rm H},r)=-1
\ ,
\label{H}
\ee
provided $\dot R(\tau_{\rm H},x)>-1$ for $x>r$ (at least locally).
From the equation of motion~(\ref{R}), this is equivalent to
$R(\tau_{\rm H},r)=2\,M(\tau_{\rm H},r)$, which
means that surfaces of different comoving coordinate $r$
reach the null surface (horizon) at different proper times.
We can equivalently define $r=r_{\rm H}(\tau)$ as the value of $r$
at which the horizon is formed at the time $\tau$.
\par
The function $r_{\rm H}(\tau)$ is of course model dependent and
affects how the effective mass evaluated on the horizon,
$M_{\rm H}(\tau)\equiv M(\tau,r_{\rm H}(\tau))$, changes in time.
In fact, its total time derivative contains two contributions,
\be
\frac{\d M_{\rm H}}{\d\tau}=\dot M_{\rm H}+M_{\rm H}'\,\dot r_{\rm H}
\ .
\label{dMHtot}
\ee
The first term in the r.h.s.~originates from the intrinsic time
dependence of the BW effective mass at constant $r$ which we have
studied in the previous Section, is a first order effect in $\epsilon$ and
would vanish in GR.
The second term accounts for the mass change due to the (possibly)
variable number of shells included within the horizon and depends on
the detailed form of the atmosphere.
Since we have assumed that our model is OS to zeroth order
in $\epsilon$, we have $M'_{\rm H}=O(\epsilon)$ outside the modified
OS boundary ($r>r_{\rm s}\sim r_0$).
\subsubsection{In the core}
\label{Hcore}
Since in this region the model is OS, the velocity $|\dot R(\tau,r)|$
increases monotonically in $r$ at fixed $\tau$.
There is therefore only an (apparent) horizon at the boundary $r=r_0$
when $R_0(\tau)$ satisfies Eq.~(\ref{H})~\footnote{We recall that
the star is now extended to spatial infinity.}.
This occurs at the time (to zeroth order in $\epsilon$)
\be
\tau_{\rm H}^{\rm OS}\equiv T-\frac{4}{3}\,M_{\rm S}
\ ,
\label{tHOS}
\ee
where $T$ fixes the time scale of the collapse (this would be the
time at which the star hits the central singularity in GR; see
also Appendix~\ref{AppDelta}).
On the event horizon of region~I, we then get
\be
\frac{\d M_{\rm H}}{\d\tau}=
\dot M_{\rm H}\equiv \dot M_0(\tau_{\rm H}^{\rm OS})
\simeq
-\frac{9\,\left(\mu-1\right)\,\epsilon}{128\,\pi\,\rho_0\,M_{\rm H}^2}
\ ,
\label{MRI}
\ee
which is precisely the Hawking flux obtained in Section~\ref{holo}
once we replace the definition $\epsilon=\rho_0\,\lambda^{-1}$.
\subsubsection{In the transition region}
\label{Htrans}
Inside this Tolman region, the horizon for a given shell will be reached
at proper time $\tau>\tau_{\rm H}^{\rm OS}$.
The partial time derivative of the effective mass on the horizon will then
scale according to Eq.~(\ref{mII}).
In particular, on considering again that $r-r_0\simeq O(\epsilon)$,
the total derivative scales as
\be
\frac{\d M_{\rm H}}{\d\tau}\simeq\dot M_{\rm H}(\tau,r_{\rm H}(\tau))
\simeq
\dot M_0(\tau)
\simeq
\dot M_0(\tau_{\rm H}^{\rm OS})
\ ,
\ee
in which the last approximate equality follows from the
condition~(\ref{R=R0}).
This implies that the flux at the OS horizon will continue up to
the time when $r_{\rm H}(\tau)=r_{\rm s}$.
\subsubsection{In the tail}
\label{Htail}
This is the most interesting part, since the above results
for the transition region allow us to approximate the boundary
of the modified OS star as the sphere $r=r_{\rm s}$.
\par
\begin{figure}[t]
\centering
{\raisebox{2.3cm}{$\frac{\d M}{\d\tau}$}}
\epsfxsize=2.8in
\epsfbox{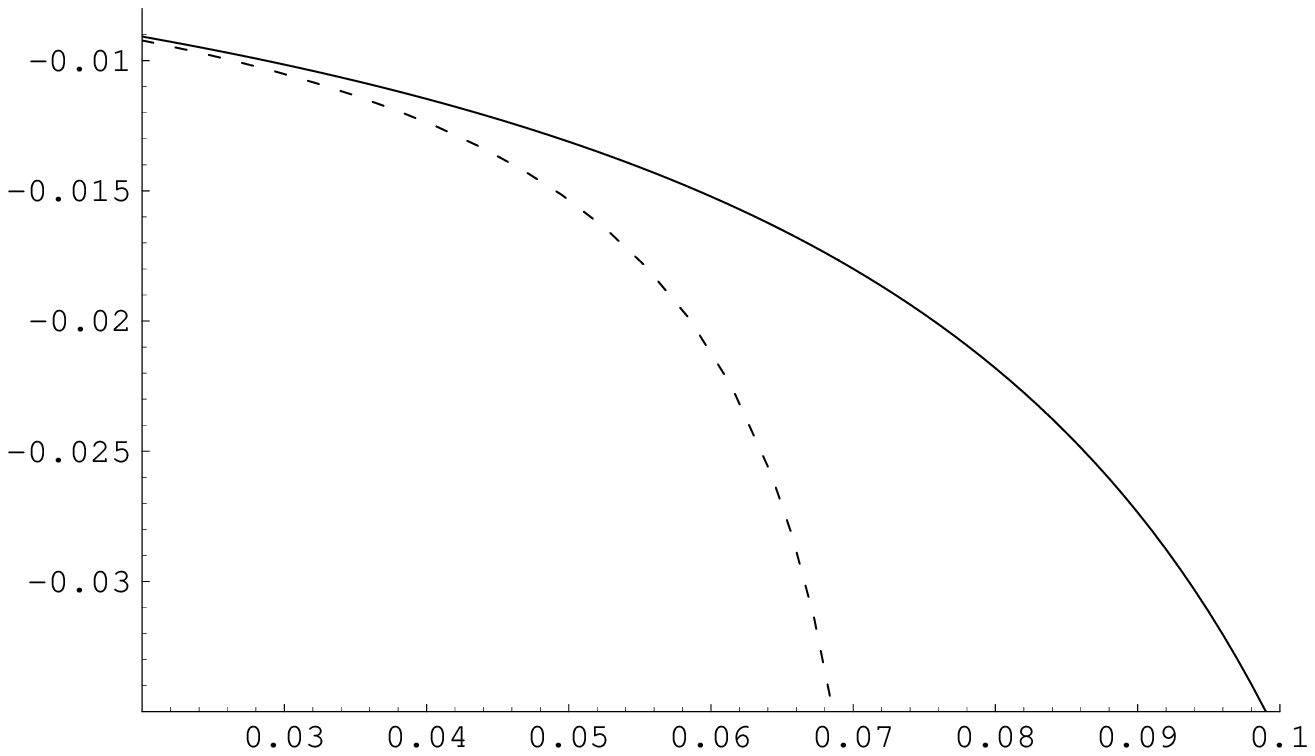}
\\
$\ \ \ \ \ \ \ \ \ \ \ \ \ \ \ \ \ \ \ \ \ \ \ \ \ \ \ \
\ \ \ \ \ \ \ \ \ \ \ \ \ \ \ \ \ \ \ \ \ \ \ \ \ \ \ \ \ \ t$
\caption{Total time derivative of the effective mass $M_{\rm H}(\tau)$
versus the time $t=(\tau-\tau_{\rm H}^{\rm OS})/T$ for the BW model
(solid line) and for the Hawking law (dotted line)
for $\epsilon=10^{-4}$, $M_{\rm S}=\rho_0=1$, $T=10$ and
$\mu=5000>4824=\mu_{\rm c}$.
\label{dMH0}}
\end{figure}

Since both $\dot M_{\rm H}$ and $M_{\rm H}'$ in Eq.~(\ref{dMHtot})
are of order $\epsilon$, we can use the zeroth order
Eq.~(\ref{motion0}) in order to determine the evolution of
the horizon which therefore stays at
\be
R(\tau,r_{\rm H}(\tau))\simeq 2\,M_{\rm S}
\ ,
\label{Hfix}
\ee
or $r_{\rm H}(\tau)\simeq r_0+\tau-\tau_{\rm H}^{\rm OS}$.
Upon inserting Eq.~(\ref{MIII}) into Eq.~(\ref{dMHtot}) with
$\dot r_{\rm H}\simeq 1$, we then find
\be
\frac{\d M_{\rm H}}{\d\tau}&\simeq&
\dot M_0(\tau)+\dot m_\cu(\tau,r_{\rm H};r_{\rm s})
\nonumber
\\
&&
+m_\rho'(r_{\rm H};r_{\rm s})
+ m_\cu'(\tau,r_{\rm H};r_{\rm s})
\ ,
\label{dMdt}
\ee
for $\tau>\tau_{\rm H}^{\rm OS}$.
As opposed to the other terms in Eq.~(\ref{dMdt}), the contribution
given by $M_{\rm H}'$ depends on the specific profile chosen for
$\rho$, and is present in GR Tolman model as well.
The increase of the mass at the horizon induced by this term is
simply due to a flux of matter flowing towards the center of the
star which makes the apparent horizon grow.
Obviously $m_\rho'$ is positive, does not explicitly depend on time
[but just via $r_{\rm H}=r_{\rm H}(\tau)$] and decreases for
increasing $r_{\rm H}$ (or, equivalently, for increasing time
$\tau$).
The deviation of the smooth energy density of the atmosphere
from the OS outer vacuum should be local, hence very much
concentrated near the OS boundary.
This implies that the profile of the density should decay very
fast~\footnote{One can for example take a Fermi distribution for
$m_\rho'$ so that it decays exponentially outside the transition
region.
We anyway note that, since $m_\rho'$ contributes positively
to $\d M_{\rm H}/\d\tau$, if it were not negligible, it would
shift the curve in Fig.~\ref{dMH0} upwards, thus making
the energy flux of a BW star differ more from the Hawking
behavior after the horizon has formed.}.
With this in mind, one can choose $r_{\rm s}$ in such a way that
$\rho\simeq O(\epsilon^2)$ and
$r_0-r_{\rm s}\simeq O(\epsilon)$, so that $m_\rho'$ is negligible
to first order in $\epsilon$.
We will then not consider its contribution to the total variation of
the mass at this stage.
The term $m_\cu'$ is instead determined uniquely by
Eq.~(\ref{dmeq}) and the initial condition for $\cu$ (which we
naturally took as zero Weyl energy).
\par
From Eqs.~(\ref{dmeq}) and (\ref{Hfix}), the first contribution is
easily approximated as
\be
\dot M_{\rm H}&\simeq&
\dot M_0(\tau)\,\left(\frac{R_0(\tau)}{2\,M_{\rm S}}\right)^{3/2}
\!\!\!\!\!
-\frac{m_\cu(\tau,r_{\rm H};r_{\rm s})}{M_{\rm S}}
\nonumber
\\
&\simeq&
-\frac{9\,\epsilon}{64\,\pi\,\rho_0\,M_{\rm S}}\,
\left[\mu-\frac{4\,M_{\rm S}^2}{R_0^2(\tau)}\right]\,
\frac{1}{R_0(\tau)}
\nonumber
\\
&&
-\frac{m_\cu(\tau,r_0+\tau-\tau_{\rm H}^{\rm OS};r_{\rm s})}
{M_{\rm S}}
\ ,
\label{dM_1}
\ee
in which we finally used Eq.~(\ref{dm0finale}) and $R_0(\tau)$
is of the form (\ref{Rmot0}) with $f(r)=0$.
\par
\begin{figure}[t]
\centering
{\raisebox{0.3cm}{$\Delta\!\!\!\!$}}
\epsfxsize=2.8in
\epsfbox{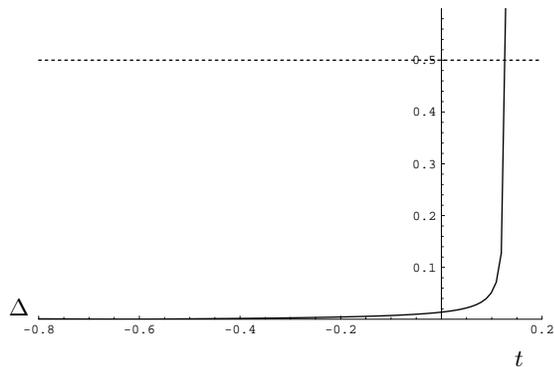}
\\
$\ \ \ \ \ \ \ \ \ \ \ \ \ \ \ \ \ \ \ \ \ \ \ \ \ \ \ \
\ \ \ \ \ \ \ \ \ \ \ \ \ \ \ \ \ \ \ \ \ \ \ \ \ \ \ \ \ \ t$
\caption{Error (\ref{errorF}) for the solid line in Fig.~\ref{dMH0}.
Note that $t<0$ correspond to times before the horizon
formation ($\tau<\tau_{\rm H}^{\rm OS}$).
The dashed line marks the limit of validity $\Delta\lesssim 0.5$.
\label{D}}
\end{figure}
Since the analytic expression for $m_\cu$ is too complicated
to display, we compute the total time derivative of the effective
mass at the horizon from the solutions of Eq.~(\ref{dmeq})
numerically.
For the values of the parameters used in Figs.~\ref{U} and
\ref{dMext0}, the result is plotted in Fig.~\ref{dMH0} up to the
time when the error estimate $\Delta\simeq 0.5$ (see Fig.~\ref{D}).
We can see that the flux is smaller with respect to that predicted
by Hawking, and this behavior remains for different values of
the parameters.
In particular, decreasing $\epsilon$ or $\mu$, as well as increasing
$M_{\rm S}$, reduces the luminosity, as expected, and keeps our
approximation reliable for longer times.
\par
The conclusion is that, although the evaporation sets out
according to Hawking's law, the back-reaction on the brane
metric subsequently reduces the emission until the effective
mass of the core vanishes and its radius bounces back
with $\dot M_0$ that becomes positive~\footnote{We recall that
$\dot M\propto \dot R$.}.
There will be an interval of time during which the error $\Delta$
is large and our first order analysis outside of the core breaks
down.
However, after a finite amount of proper time, $\Delta$
will become small again and the system will evolve back to the
initial condition through a sequence of states obtained by
inverting the time in the above solution.
\subsection{Luminosity}
A distant observer experiences an impinging flux of energy
during the collapse, whose total amount must be calculated
from the horizon to infinity (since the region
inside the horizon is causally disconnected from such
an observer).
Since the total energy from the origin to infinity is conserved
and the process extracts energy from the hole, we expect that
the measured flux is positive.
\par
After the horizon has formed on the boundary of the star
(more explicitly, for $r_{\rm H}\geq r_{\rm s}\sim r_0$ and
$\tau\gtrsim\tau_{\rm H}^{\rm OS}$),
one has
\be
\Phi_\tau&\equiv&
\frac{\d}{\d\tau}\left[\lim_{\bar r\to\infty}
\frac{4\,\pi}{3}\,\int^{\bar r}_{r_{\rm H}(\tau)}
\rho^{\rm eff}\,\left(R^3\right)'\,\d r\right]
\nonumber
\\
&=&
\lim_{r\to\infty}\dot M(\tau,r)
-\frac{\d M_{\rm H}(\tau)}{\d\tau}
=
- \frac{\d M_{\rm H}(\tau)}{\d\tau}
\ ,
\ee
in which we used the conservation of the total effective
mass~(\ref{dMinfty}).
Further, since $\partial_\tau\to\partial_t$ for $r\to\infty$ one
finally obtains the luminosity
\be
\Phi_t
\simeq -\frac{\d M_{\rm H}}{\d\tau}
\ .
\label{comp}
\ee
The flux seen by a distant observer therefore shows the same
dependence on the mass $M_{\rm H}$ as the semiclassical expression
when the horizon is first forming, and subsequently decreases to zero
(before it becomes negative).
However, since this happens after the apparent horizon begins to
form, a distant observer might have to wait an infinite amount
of time to measure a vanishing flux.
\par
We now consider the only instant when the Hawking radiation
actually equals the BW result, that is at the OS boundary
when $r_{\rm H}(\tau)=r_0$.
Reintroducing the Newton's constant $G$ in units $c=\hbar=1$
(so that $M$ has the dimension of a mass in this Section,
and not of a length) and the definition of $\epsilon$,
we obtain
\be
\Phi_t
\simeq \frac{9\,(\mu-1)}{128\,\pi\,G^4\,\lambda\,M_{\rm H}^2}
\ ,
\label{our}
\ee
which we can compare with the semiclassical luminosity as
calculated in the Schwarzschild background~\cite{Page}
\be
\Phi_t\simeq \frac{\alpha}{G^2\, M_{\rm H}^2}
\ ,
\label{emp}
\ee
where $\alpha$ is a dimensionless coefficient which depend
on the quantum field theory chosen.
\par
An astrophysical object has $M\gg 10^{31}\, {\rm GeV}$ and
we can therefore use the result~(\ref{emp}) of Ref.~\cite{Page},
that is $\alpha=\alpha_0\,N^2\simeq 7.74\cdot 10^{-3}\,N^2$,
where $N$ is the number of particle species appearing in the
quantum theory~\footnote{In particular, for this range of masses,
it was proven in Ref.~\cite{Page} that only massless spin $0$, $1$
and $1/2$ particles contribute.}.
On now using the lower ($\mu_{\rm c}$) and upper
($\mu_{\rm a}\sim\mu_{\rm c}^{3/2}$)
bounds for $\mu$ as given in Eqs.~(\ref{2_0}) and~(\ref{upper_mu}),
we obtain a limit on the number of species that can take part
in the Hawking process,
\be
\frac{9}{128\,\pi\,\alpha_0}\,\left(\frac{32}{3}\,\pi\right)^{2/3}\,
\left(\frac{M^4}{\lambda}\right)^{1/3}
<N^2\ll\frac{G\,M^2}{2\,\alpha_0}
\ .
\ee
The result of Ref.~\cite{Page} must be valid for any mass
$M\gg 10^{31}\,$GeV
Therefore, for the upper bound
we can safely consider $M\sim 10^{35}\,$GeV, so that
\be
10^{43}\,\left(\frac{{\rm GeV}^4}{\lambda}\right)^{1/3}\ll
N^2\ll 10^{33}
\ .
\ee
For consistency, we must also have
\be
\lambda\gg 10^{30}\, {\rm GeV}^4
\ .
\ee
Considering that the AdS length $\ell=\sqrt{3/4\pi\, G\,\lambda}$
must be much larger than the Planck length, we finally obtain
\be
10^{-32}\, {\rm mm}\ll \ell \ll 10^{-9}\, {\rm mm}
\ ,
\label{sb}
\ee
This bound for the AdS length is three orders of magnitude better
than the best constraint found in Ref.~\cite{emparan} considering
the time scale of primordial black hole evaporations.
\subsection{Trace anomaly}
Strictly speaking, there is no trace anomaly in our approach,
since we have included the back-reaction of the effective matter
on the brane metric.
However, in order to compare with known results {\em without\/}
the back-reaction, we can define the trace anomaly ${\mathcal R}$
as the sum of the Ricci scalar and the trace of the {\em bare\/}
stress tensor~\footnote{We recall that $R^\mu{}_\mu=-8\,\pi\, T$
in GR, whereas $R^\mu{}_\mu=-8\,\pi\, T^{\rm eff}$ in the BW.}.
From the effective Einstein equation~(\ref{Ricci}) one readily
obtains (see also Ref.~\cite{shiroida})
\be
{\mathcal R}&\equiv&
R^\mu_{\ \mu}+8\,\pi\,T^\mu{}_\mu
=
-8\,\pi\,\frac{\rho^2}{\lambda}
\nonumber
\\
&=&
-\frac{1}{2\,\pi\,\lambda}\,
\left(\frac{M_0'}{R^2\,R'}\right)^2
\ .
\label{badR}
\ee
At the OS boundary, $r=r_0$, we then have
\be
{\mathcal R}=
-\frac{9}{2\,\pi\,\lambda}\,\frac{M_{\rm S}^2}{R^6}
\ ,
\ee
which is the quantum Ricci anomaly of Ref.~\cite{CF} with
the correct sign at the collapsing boundary.
It is then clear that the sign mismatch found in Ref.~\cite{BGM}
was due to the choice of a non-smooth energy density and that,
by adding a tail, we have described how the excess energy stored
in the OS boundary is released.
\par
For $r>r_0$, we have ${\mathcal R}=O(\epsilon^{2})$, which is
therefore negligible from the point of view of our analysis because,
as we have shown, the Hawking flux diverges in time whereas the BW
one remains finite outside the OS boundary.
How the back-reaction on the brane metric gradually annihilates
the Ricci anomaly in the transition region can only be understood
by introducing specific models for the tail which must also be
consistent with the five-dimensional problem,
and goes beyond the investigation we want to present here.
In any case, if the holographic analogy holds, the anomaly of
Ref.~\cite{CF} must just be effective at the boundary of the OS
core and decrease to zero at the modified boundary of the star
($r=r_{\rm s}$).
It is so because the back-reaction on the brane metric must be
consistent with the modified Einstein equations, whereas in
Ref.~\cite{CF} the Einstein equations are just solved for the
background.
\section{Dissipative collapse}
\label{heat}
We now wish to discuss the possible effects of a dissipative
term of the form $Q_a\neq 0$ in our model, although including
such an energy flow would require heavy numerical investigations
of the full five-dimensional equations and goes beyond the
scope of the present paper.
\par
Since, the OS core of our dust star is non-dissipative
by construction, the off-diagonal term $V(\tau,r)=0$
for $r<r_0$ in the bulk metric (\ref{bulk_m}), as we
explained in Section~\ref{ssd}.
This implies that one can have $Q_a(\tau,r)\sim V(\tau,r)\neq 0$
only for $r>r_0$ with an OS-like core.
In this case, compatibility with the Hawking effect
would constrain the heat to flow from the OS boundary
towards infinity, and the energy of the transition region
and tail would thus be dissipated away completely after
a suitable amount of time.
In the meanwhile, the core should keep bouncing back and forth
between its initial condition and the state with vanishing
effective mass, since its evolution cannot be affected by
$Q_a$ in the atmosphere.
The net final result should thus be that the system converges
to the model with an empty exterior discussed in Ref.~\cite{BGM},
which we already know is not acceptable in the BW.
This argument shows that a non-dissipative core is most likely
incompatible with a heat flow in the external region and
that the condition $Q_a=0$ should therefore not represent
a real restriction for the model we have analyzed.
\par
Of course, a more realistic model for a collapsing star
should also have a dissipative core and one should consider
a non-vanishing $Q_a$ everywhere, as well as a non-vanishing
flow of matter.
The (absolute value of the) total (holographic) flux of energy
measured far away from the core would then be larger than the
one from a non-dissipative core, and therefore closer in value
to the four-dimensional Hawking flux.
We nonetheless expect that a global horizon does not
form, since the total outgoing flow will make the star
``evaporate'' until all the initial energy has been
radiated away.
This can happen either before or after the bouncing, which
does no more allow the star to come back to its initial
condition because of dissipation.
In fact, we expect that no singularity forms even in the
general case because the badly diverging part of the Ricci
scalar proportional to the squared energy density in
Eq.~(\ref{badR}), which arises from the junction conditions
on the brane, cannot be canceled by the Weyl contribution.
The singularity must then be avoided either with the help
of the Weyl tensor or by means of severe modifications to the
matter profile due to BW effects.
In the former case, the star will still bounce, whereas in the
latter it will completely ``evaporate'' before reaching
the singularity.
This anyways remains an open question that cannot be addressed
here.
\section{Conclusions}
\label{conc}
Inspired by the conjecture that classical black holes in the BW may
reproduce the semiclassical behavior of four-dimensional black holes,
we have studied the gravitational collapse of a spherical star of dust
in the RS scenario in order to clarify the underlying dynamics that
leads to this interpretation.
Regularity of the bulk geometry requires continuity of the matter
stress tensor on the brane and can lead to a loss of mass from
the boundary of the star.
We have in particular shown that, excluding energy fluxes coming
from the bulk Weyl tensor, a collapsing spherical star must
have a spatially anisotropic, although isotropic in the angular
directions, atmosphere, in order to have asymptotically flat solutions.
Interestingly, such a feature is also present in the stress tensor of
quantum fields on the Schwarzschild background~\cite{CF}.
\par
We found that the system of effective BW equations is closed to
our level of approximation and leads to the collapsing dust star
emitting a flux of energy which, at relatively low energies,
approaches the Hawking behavior when the (apparent) horizon is being
formed (let us note that similar features seem to appear for a quantum
black hole~\cite{bow} as well as in the semiclassical treatment of
collapsing shells~\cite{acvv}).
Although we cannot determine a precise value for such a flux,
which depends on the strength of the dark energy $\cu\sim\mu$,
consistency of the model constrains $\mu$ for astrophysical
objects both below and above.
With that, we were able to suggest the new stronger bound
(\ref{sb}) for the brane tension by comparing our results with
standard four-dimensional quantum computations of the Hawking
flux for astrophysical objects.
Further, inside the star $\cu$ is negative, so that each dust shell
mostly releases energy into the next shell of larger radius and the
whole process occurs mainly on the brane.
This behavior then changes gradually moving to the exterior of
the star, where $\cu$ becomes positive and the energy lost from
the core is mainly converted into bulk gravitational waves.
\par
We have also shown that the collapsing core will reach a minimum
after a finite proper time and the collapse will then turn into an
explosion which drives the whole system back to the initial state.
This happens because the BW correction to the matter stress
tensor acts as an ``anti-evaporating'' contribution which becomes
bigger as the energy increases.
The bounce will occur after the formation of the apparent
horizon (so that a distant observer presumably experiences the
explosion only after a very long amount of time) and {\em will not
allow the formation of a global event horizon\/}.
Interestingly, from the Quantum Gravity side, it seems that a similar
scenario would solve the information loss problem~\cite{Haw}.
In fact, such a behavior for the core was previously obtained in an
improved semiclassical treatment of the OS model in Ref.~\cite{impBO},
where quantum gravitational fluctuations were shown to have
effects like those which the Weyl term causes in the present context.
In any case, one might reasonably question that the bouncing ends
back to the {\em exact\/} initial state.
Let us then remark that matter in the OS model is frictionless dust,
and that, in a more realistic case, friction would of course
dissipate energy and make the evolution irreversible, as we
discussed in Section~\ref{heat}.
\par
The trace anomaly of four-dimensional quantum field theory on
the Schwarzschild background has also been naturally interpreted
as the BW correction to the trace of the matter stress tensor at
the boundary of the core.
Moreover, it has been shown that the back-reaction on the brane
metric effectively annihilates the anomaly throughout the
transition region and into the tail, compatibly with the effective
four-dimensional Einstein equations, unlike semiclassical
computations in which the Einstein equations are solved at the
purely classical level (zero order in the Planck constant).
Thus, if one believes in the holographic interpretation, it seems
that the quantum anomaly would disappear to first order in the
Planck constant when properly considering the back-reaction on
the metric.
\par
Let us finally point out that all the above features were obtained
for black holes formed by gravitational collapse, excluding therefore
primordial black holes about which we have nothing to say.
\acknowledgments
The authors would like to thank Akihiro~Ishibashi,
Christophe~Galfard and Misao Sasaki for useful comments and discussions.
C.G.~would like to thank Roy~Maartens for making him
interested in the topic and Roy~Maartens and Carlos~Barcelo
for illuminating discussions.
C.G.~would also like to thank the University of Portsmouth
ICG group for partially supporting this work and the Physics
Department of the University of Bologna for the hospitality
during part of this research.
C.G.~is supported by PPARC research grant PPA/P/S/2002/00208.
\par\noindent
\appendix
\section{Diverging effective mass in a Tolman core}
\label{App_T}
We shall here show that a general Tolman metric should have
a bounce as well as an "anti-evaporating'' phase in the BW.
We shall just consider cases in which the space-time is globally
hyperbolic and has a non-compact Cauchy surface
(as it seems reasonable for a physical gravitational
collapse).
\par
Assume that the Weyl tensor is zero.
Given a null vector $k^\mu$, from the effective Einstein
equations~(\ref{eq:effective}) we have
\be
R_{\mu\nu}\,k^\mu\,k^\nu
&=&T_{\mu\nu}\,k^\mu\,k^\nu
=(\rho^{\rm eff}+p^{\rm eff})\,\left(u_\mu\,k^\mu\right)^2
\nonumber
\\
&=&\left(\rho+\frac{\rho^2}{\lambda}\right)\,
\left(u_\mu\, k^\mu\right)^2
\geq 0
\ .
\ee
This condition ensures that the singularity theorem of
Ref.~\cite{Penrose} holds and the space-time will therefore
reach a singular point at finite proper time.
In particular, for the Tolman geometry, this means that $R\to 0$
after a finite amount of proper time and the total effective mass
in Eq.~(\ref{Meffg}) with $\cu=0$,
\be
M=m_\rho+\frac{2\,\pi}{3\,\lambda}\,\int_0^r
\rho^{2}\,\left(R^{3}\right)'\,\d x
\equiv m_\rho+\Delta M
\ ,
\label{DM}
\ee
will correspondingly diverge, as we shall now show
explicitly.
\par
First of all, let us note that the equation of motion
for dust shells inside the core ($0\le r<r_0$) yield
\be
R^2\,\dot R^2&=&
2\,m_\rho(r)+2\,\Delta M(\tau,r)
\ge
2\,m_\rho(r)
\ ,
\ee
since $\Delta M(\tau,r)>0$.
On considering that the flat Tolman solution~\cite{tolman}
satisfies the equation
\be
R^2_{\rm T}\,\dot R_{\rm T}^2=2\,m_\rho(r)
\ ,
\ee
we then have
\be
R(\tau,r)\leq R_{\rm T}(\tau,r)
\ ,
\ee
for collapsing solutions with $\dot R<0$ and
$\dot R_{\rm T}<0$.
By re-scaling the coordinate $r$,
one can always write~\cite{LL}
\begin{subequations}
\be
R_{\rm T}(\tau,r)=g(r)\,\left[\tau_0(r)-\tau\right]^{2/3}
\ ,
\label{R0}
\ee
with $g(r)$ as in Eq.~(\ref{g(r)}), and
the corresponding bare mass is given by
\be
m_{\rho}(r)=M_{\rm S}\,\left(\frac{r}{r_0}\right)^3
\ ,
\label{M0}
\ee
\end{subequations}
where $M_{\rm S}=m_\rho(r_0)$.
The function $\tau_0(r)$ represents the (proper) time at which
the shell of comoving radius $r$ hits $R_{\rm T}=0$ in GR and
must be monotonically non-decreasing in $r$ in order to avoid
shell crossings (for the OS case, one has $\tau_0=T$ independently
of $r$).
The BW correction to the effective mass can thus be estimated
as
\be
\Delta M=
\frac{1}{8\,\pi\,\lambda}\,\int_0^r
\frac{\left(m_\rho'\right)^2}{R^{2}\,R'}\,\d x
\geq
\frac{1}{8\,\pi\,\lambda}\,\int_0^r
\frac{\left(m_\rho'\right)^2}{R_{\rm T}^{2}\,R_{\rm T}'}\,\d x
\ .
\label{D_M}
\ee
The last integral is not well-defined for all $\tau>0$.
In fact, $R_{\rm T}\to 0$ for $\tau\to\tau_0(r)$ and $\Delta M$
(which is proportional to $\rho^{2}$) therefore diverges
for $r\to\tau_0^{-1}(\tau)$.
To see this more clearly, let us define
$T_\varepsilon\equiv\tau_0(\varepsilon)$
the time at which a shell infinitesimally close to the center
(at $r=\varepsilon$ with $0<\varepsilon\ll r_0$)
hits the singularity, $R_{\rm T}(T_\varepsilon,\varepsilon)=0$.
For $\tau\ge T_\varepsilon$, one can formally split the last
integral in Eq.~(\ref{D_M}) into two parts (we include
irrelevant numerical factors in $\zeta\neq 0$ and finite),
\be
\Delta M\geq
\zeta\int_0^{\tau_0^{-1}(\tau)}
\frac{\left(m_\rho'\right)^2}{R_{\rm T}^{2}\,R_{\rm T}'}\,\d x
+\zeta\int_{\tau_0^{-1}(\tau)}^r
\frac{\left(m_\rho'\right)^2}{R_{\rm T}^{2}\,R_{\rm T}'}\,\d x
\ .
\label{2int}
\ee
The first integration is over matter already collapsed into
the point-like singularity (whose volume element
$R_{\rm T}^2\,R_{\rm T}'=0$) and the corresponding $r$
does not represent a valid spatial coordinate any longer.
One might try to regularize this integral.
However, the second integration, which is instead over a valid
range of the coordinate $r$, diverges~\footnote{One would also
have a divergence for $\tau=\tau_0(r)+(2/3)\,r\,\tau_0'(r)$,
but since $\tau_0'(r)>0$, the latter occurs at later times and
we have ignored it.} and leads to
\be
\Delta M
&\gtrsim&
\int_{\tau_0^{-1}(\tau)}^r
\frac{x^2\,\d x}{\left[\tau_0(x)-\tau\right]\,
\left[3\,\tau_0(x)+2\,x\,\tau_0'(x)-3\,\tau\right]}
\nonumber
\\
&\sim&
\int_0^{\tau_0(r)-\tau}\frac {\d y}{y}
\sim
\infty
\ .
\label{div}
\ee
\par
\begin{figure}[t]
\centering
{\raisebox{2.2cm}{$\Delta$}}
\epsfxsize=2.8in
\epsfbox{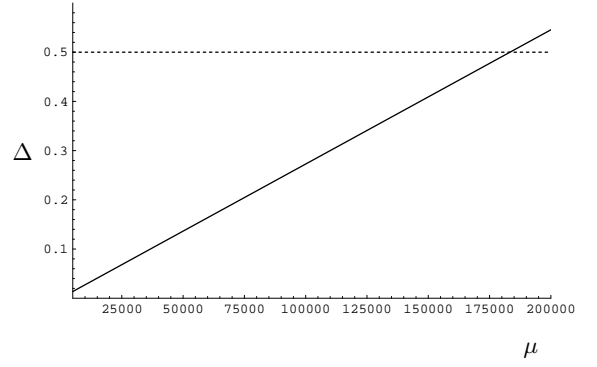}
\\
$\ \ \ \ \ \ \ \ \ \ \ \ \ \ \ \ \ \ \ \ \ \ \ \ \ \ \ \
\ \ \ \ \ \ \ \ \ \ \ \ \ \ \ \ \ \ \ \ \ \ \ \ \ \ \ \ \ \ \mu$
\caption{Error (\ref{errorF}) evaluated on the horizon at
$\tau=\tau_{\rm H}^{\rm OS}$ for $M_{\rm S}=\rho_0=1$,
$\epsilon=10^{-4}$, $T=10$ and $\mu>\mu_{\rm c}=4824$.
The dashed line marks the limit of validity $\Delta\lesssim 0.5$.
\label{errorMu}}
\end{figure}
\begin{figure}[t]
\centering
{\raisebox{2.2cm}{$\Delta$}}
\epsfxsize=2.8in
\epsfbox{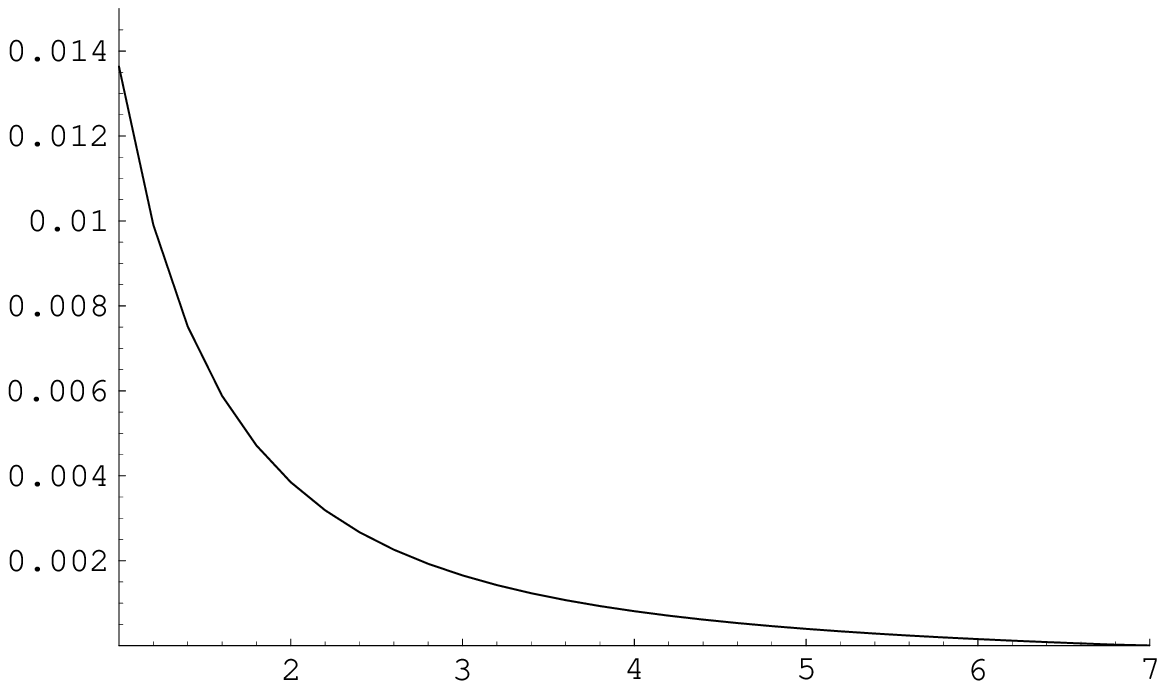}
\\
$\ \ \ \ \ \ \ \ \ \ \ \ \ \ \ \ \ \ \ \ \ \ \ \ \ \ \ \
\ \ \ \ \ \ \ \ \ \ \ \ \ \ \ \ \ \ \ \ \ \ \ \ \ \ \ \ \ \ M_{\rm S}$
\caption{Error (\ref{errorF}) evaluated on the horizon at
$\tau=\tau_{\rm H}^{\rm OS}$ for $\rho_0=1$,
$\epsilon=010^{-4}$, $T=10$ and
$\mu=5000>4824=\mu_{\rm c}(M_{\rm S}=1)$.
\label{errorMs}}
\end{figure}
The only way to avoid the above divergence is to have a negative
Weyl tensor.
If $\cu\neq -\rho^2/2\lambda$, the singularity theorem is violated
and the collapse will experience a bounce as discussed in the text,
whereas for $\cu= -\rho^2/2\lambda$ the NLCE's imply that
\be
\Pi=
\frac{6}{\lambda\, R^3}\,\int^r_0 \rho^2\,R^2\,R'\,
\d x+\ldots
\ ,
\ee
where the dots stand for harmless terms.
As we have shown, the above expression diverges as soon as any
shell at $r=\varepsilon\ll r_0$ approaches the singularity,
thus making the anisotropic stress tensor singular in an extended
region $r>0$.
If one requires that the four-dimensional space-time is regular
everywhere and at any time, a part from isolated points, one should
then consider $\cu<0$ and
sufficiently large (with $\cu\neq -\rho^2/2\lambda$), so as to make
the collapse bounce in the Tolman case as well.
\par
In light of the above analysis, we believe that the bouncing is
a general feature of the gravitational collapse in the BW,
although a numerical analysis of the five-dimensional equations is
needed to ensure the absence of other singularities in the bulk.
\par
\begin{figure}[ht]
\centering
{\raisebox{2.2cm}{$\Delta$}}
\epsfxsize=2.8in
\epsfbox{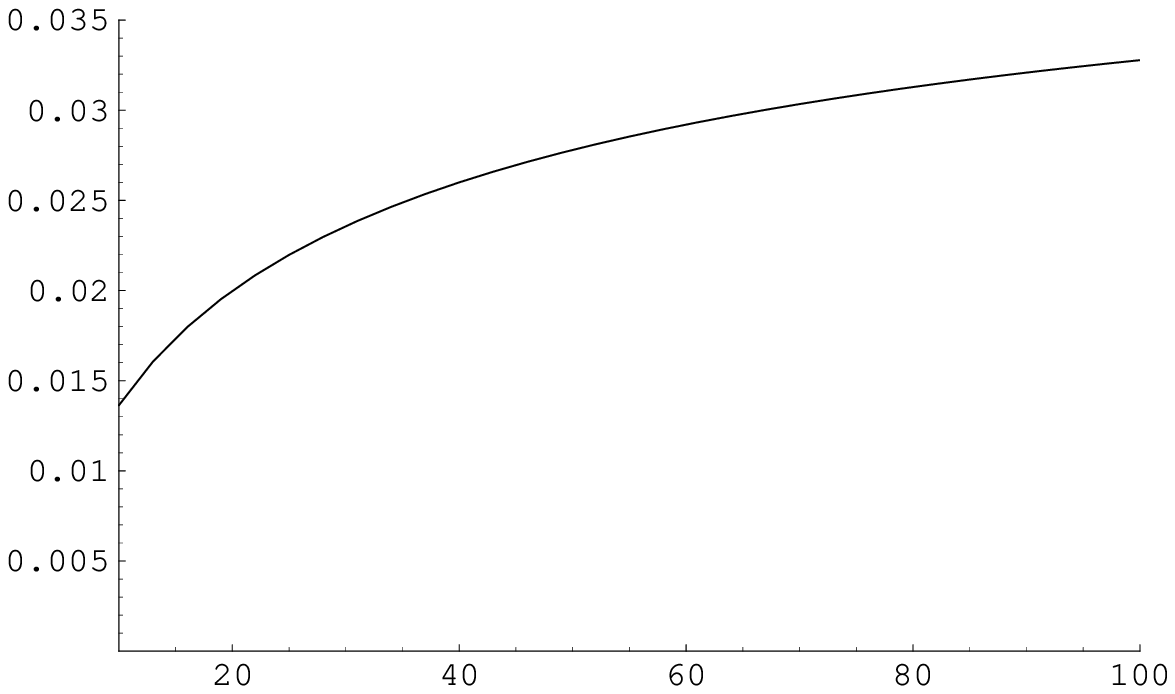}
\\
$\ \ \ \ \ \ \ \ \ \ \ \ \ \ \ \ \ \ \ \ \ \ \ \ \ \ \ \
\ \ \ \ \ \ \ \ \ \ \ \ \ \ \ \ \ \ \ \ \ \ \ \ \ \ \ \ \ \ T$
\caption{Error (\ref{errorF}) evaluated on the horizon at
$\tau=\tau_{\rm H}^{\rm OS}$ for $M_{\rm S}=\rho_0=1$,
$\epsilon=10^{-4}$, and $\mu=5000>4824=\mu_{\rm c}$.
\label{errorT}}
\end{figure}
\section{Error estimates}
\label{AppDelta}
We shall here display a few plots to clarify the behavior of the
function $\Delta$ defined in Eq.~(\ref{errorF}) as an estimate
of the error produced by truncating to first order in $\epsilon$.
In particular, it is clear from Fig.~\ref{errorMu} that $\Delta$ grows
linearly with $\mu$ and, from Fig.~\ref{errorMs}, that it instead decreases
with increasing $M_{\rm S}$.
\par
We finally show in Fig.~\ref{errorT} that the approximation on the
horizon becomes worse for increasing $T$, the proper time at which the
OS core would hit the central singularity in GR.
This parameter has no physical meaning for the bouncing
core, hence can be fixed by minimizing the error $\Delta$ in the
time interval of interest.
Note, however, that we need $T>4\,M_{\rm S}/3$ in order to have
$\tau_{\rm H}^{\rm OS}>0$, and we also want that the horizon forms
a relatively long time after the system begins to evolve.
We therefore start this graph from $T=10$, which is the same value
we use as a fair optimization for the quantities plotted in
Figs.~\ref{U}-\ref{dMH0}.
%
%
%

%
\end{document}